\documentclass[12pt]{article}
\usepackage[utf8]{inputenc}
\usepackage{setspace}  
\usepackage[margin=1in]{geometry}

\usepackage{booktabs}
\usepackage{longtable}
\usepackage{comment}
\usepackage[hyphens]{url} % not crucial - just used below for the URL

\usepackage{natbib}
\usepackage{float}
\usepackage{graphicx}
\usepackage{listings}
\usepackage[]{algorithm2e}
\usepackage{amssymb}
\usepackage{amsmath}
\usepackage{gensymb}
\usepackage[font=footnotesize,labelfont=bf]{caption}
\usepackage{pgf, graphics}
\usepackage{multirow}
\usepackage{makecell}
\usepackage{array}
\usepackage{placeins}

\newcolumntype{P}[1]{>{\centering\arraybackslash}m{#1}}


\usepackage{amsmath}
\DeclareMathOperator*{\argmax}{arg\,max}
\DeclareMathOperator*{\argmin}{arg\,min}

\begin{document}

\title{\bf Unsupervised Methods for Identifying \\Pass Coverage Among Defensive Backs \\with NFL Player Tracking Data}
\author{Rishav Dutta$^1$, Ronald Yurko$^2$, Samuel L. Ventura$^{2,3}$}
% \author{Author Names Blinded For Peer Review}
\date{%
    \footnotesize
    $^1$Carnegie Mellon University, School of Computer Science\\%
    $^2$Carnegie Mellon University, Department of Statistics \& Data Science\\%
    $^3$Pittsburgh Penguins\\[2ex]%
    \today
}

%\date{\today}

\maketitle

%\begin{comment}
\begin{abstract}
Statistical analysis of defensive players in football has lagged behind that of offensive players, special teams, and coaching decisions, largely because data on individual defensive players has historically been lacking.  With the introduction of player tracking data from the NFL, researchers can now tackle these problems.  However, event and strategy annotations in the NFL's tracking data are limited, especially when it comes to describing what defensive players do on each play.  Moreover, methods for creating these annotations typically require extensive human labeling, which is difficult and expensive.  Because of the importance of the passing game and the limited prior research on the defensive side of passing, we provide annotations for the pass coverage types of cornerbacks using unsupervised learning techniques, which require no training data.  We define a set of features from the tracking data that distinguish between ``man'' and ``zone'' coverage.  We use mixture models to create clusters corresponding to each group, allowing us to provide probabilistic assignments to each coverage type (or cluster).  Additionally, we quantify each feature's influence in distinguishing defensive pass coverage types.  Our work makes possible several potential avenues of future NFL research into defensive backs and pass coverage strategies.
\end{abstract}
%\end{comment}

\section{Introduction}

%  Need to re-write introduction

%National Football League (NFL) teams are increasingly becoming more pass-oriented on offense.  NFL offenses are much more inclined to throw the ball than to rush. Defenses, however, have adapted to counter these strategies. Increasingly, defenses are able to force the offense into bad throws, or even prevent the quarterback from throwing at all. While all eleven defenders on the defense are involved, the primary types of players the quarterback has to account for on the defense are the defensive backs. Generally, the primary objective of the is to prevent the completion of passes, whether it be through a form of zone coverage or man-to-man coverage. 

%In 2019, the National Football Leauge (NFL) released player and ball tracking data from the first six weeks of the 2017 season for its inaugural Big Data Bowl.  This tracking data, maintained by NFL Next Gen Stats, marks the locations and trajectories (speed, angle) of all 22 players on the field and the ball at a rate of 10 Hz.  The NFL is the first major professional sports league to release such detailed and abundant tracking data to the public, ushering in an era of innovative research into the sport of American football.

%Prior to the release of this data, p
Statistical analysis of defensive players in the National Football League (NFL) is challenging, since the data collected and made publicly available for each play is limited to players directly involved in a play (e.g. the passer, rusher, receiver, tackler, etc).  However, most of what happens throughout the course of a football play happens away from the ball:  offensive linemen blocking, defensive linemen going after the quarterback or ball-carrier, defensive backs covering receivers, etc.  To date, publicly available statistics for the evaluation of defensive players are limited to simple counting measures like tackles, sacks, interceptions, etc.  Private companies like Pro Football Focus likely provide innovative metrics to NFL teams, but this information is not available publicly.  Aside from what has been recently made available via the NFL's Next Gen Stats website, little work has been done publicly to analyze defensive players.  

However, the NFL now collects detailed player and ball tracking data, which marks the locations and trajectories (speed, angle) of all 22 players on the field and the ball at a rate of 10 Hz.  Using this data, new measures of player performance are now possible for the first time.  For example, \cite{Burke18} provides a method for analyzing pass rushers using NFL tracking data.  

Across all team sports, analyzing on-field events or strategies employed by teams or individual athletes is only possible if these events or strategies are annotated in the tracking dataset.  For example, the NBA provides basic event annotations (e.g. passes, shots, turnovers) in its optical tracking data, but more detailed information (e.g. defensive schemes, picks, or set plays) must be identified by analysts.  An example of this is the work of \cite{Miller17}, who identify and annotate set plays during basketball possessions.  Similarly, the NFL provides some basic event annotations (e.g. ball snapped, handoff, first contact, etc), but more detailed annotations -- especially those relating to defensive players -- must be identified by analysts.  We describe prior work in this area below.

\subsection{Prior Research on Annotating Events and Strategies with Tracking Data in Team Sports}

%Since the NFL tracking data was released early 2019, there is not much work done involving providing annotations for tracking data in football.  As mentioned above, the work of \cite{Chu19} is a notable exception here.  However, similar work has been done in other sports. Most specifically, work with tracking data has been done in other ``invasion'' style sports, such as soccer and basketball. 

Prior researchers have attempted to annotate individual events and strategies (in both football and other sports) using both supervised and unsupervised learning techniques.  Here, it is first important to distinguish unsupervised learning techniques, which do not require potentially expensive sets of training data, from supervised learning techniques, which require training data to model specific outcomes.  If researchers have access to an extensive set of ground truth labels for the outcome they seek to model, they should use a supervised learning approach like \cite{DeepQB}, which uses training data to predict which potential pass catcher a quarterback will target, among other contributions.  Without labels, unsupervised techniques are preferable, as the researchers below have demonstrated.

In soccer, \cite{soccerstyle} use player tracking data to identify a team's playing style from their players' positions and movements on the field.  To do this, they design a set of role-specific features that reduces the entropy of role-specific occupancy maps. In doing so, the authors assign some formation-based roles to individual players, and then create occupancy maps based on these roles that are assigned. 

In basketball, \cite{openshot} use the spatio-temporal changes in the team formation in order to determine what features allow for a player to create an ``open'' shot.  Similar to \cite{soccerstyle}, the authors create role-based features by initially assigning each player a role.  In this case, they assign each of the five players on the court one of the traditional five basketball positions.  After doing this, they can track the motion of the roles rather than individual players in order to create a permutation-free set of features.  

\cite{gudmundsson1} provides an excellent overview of prior work on formation identification in team sports.  We encourage interested readers to visit this paper for more information on this topic. %Some work has been done in attempting to do this in hockey, where attempts were made to assign player roles in a one-to-one mapping throughout frames.
Most prior work in this area has focused on annotating team-wide strategies, as opposed to the strategies of individual athletes, as we do in this paper.  Two exceptions are \cite{Miller17} and \cite{Chu19}, who use unsupervised learning techniques such as mixture modeling to provide these supplemental annotations to users of the data.  As mentioned above, \cite{Miller17} annotate set plays in the NBA.  To do this, they use complex mixtures of Bezier curves to cluster the simultaneous trajectories of the five players on the court.  \cite{Chu19} use a similar approach, but for individual wide receiver routes in football.  

The problem we tackle in this paper deviates from these prior approaches in two important ways.  First, unlike most prior work, we seek to identify the coverage types of \emph{individual} defensive players, rather than look at team-wide strategies.  Second, unlike \cite{Miller17} and \cite{Chu19}, who each use mixture models to provide unsupervised labels for the patterns of movement of offensive players, we focus on defensive players.  This is important, since in most team sports, offensive players dictate their own movements, while defensive players typically move in reaction to offensive players.  Because the movements of players in defensive pass coverage are inherently reactive, we cannot simply cluster the defensive backs trajectories and expect to identify pass coverage types from the results.  Instead, we take the approach described below.

%Once event or strategy annotations are in place, further analysis into the efficacy of each strategy or the value of each event can be conducted with relative ease.  However, annotation can be challenging, often requiring a combination of subject matter expertise and advanced statistical modeling.

\subsection{Our Contribution}

In this paper, we introduce a unsupervised approach for annotating (or labeling) the coverage type of defensive backs in American football using NFL tracking data, using an extensive set of features that we create to characterize and distinguish the motions of defensive backs in relation to the corresponding offensive player.  The unsupervised approach that we use has several benefits (described below), and can be extended to other event and strategy annotation problems in NFL tracking data.  We choose to tackle the problem of identifying the type of pass coverage used by defensive backs for several reasons (described below).

First, pass coverage types are not available publicly on a play-by-play level.  NFL coaches may collect this information on their own (or turn to third-party companies like Pro Football Focus to provide this information), but doing so can be time-consuming and/or costly.  By automating this process, we allow coaches to spend their time more effectively, and we allow resources to be allocated to other organizational needs.  Importantly, our model can be used by NFL teams instantaneously during a game (assuming tracking data is available in real-time); this would, for example, allow coaches to have more information at hand when analyzing opponents' strategies, and make on-the-fly adjustments at halftime or throughout the course of the game without the need for humans to manually label each play.  Moreover, not all users of the NFL player tracking data also have access to an extensive set of plays with labeled pass coverage types.  By automating this process using unsupervised techniques, we level the playing field for users of the data who lack this information or the resources to acquire it.

Second, passing has become increasingly important in the NFL in recent years, with teams passing the ball (vs. running the ball) more than they ever have before.  Despite this, there has been limited public research into the play of defensive backs and the efficacy of different coverage schemes.  By providing coverage type annotations, we pave the way for future researchers to tackle these important problems.

Third, this problem is both challenging and adaptable to other event and strategy annotation problems.  Unlike the work of \cite{Miller17} or \cite{Chu19}, who provide annotations for offensive player movement (e.g. wide receiver routes), we address an analogous problem for defensive players.  The key difference here is that offensive players control their movements on the playing surface, while defensive players act in reaction to the offensive players\footnote{There are exceptions here, e.g. option routes where the offensive player's route is chosen during the play in reaction to the pattern of motion of the defensive back.}.  This means that techniques that cluster player trajectories on the field are not appropriate for identifying defensive coverage schemes.  Instead, we generate a rich set of features describing the movements of defensive players \emph{in relation to their teammates and to their counterparts on the offensive side} (e.g. wide receivers).  We use this extensive set of features to identify groups of plays with similar coverage, then use mixture models to assign ``soft'' cluster labels like ``man coverage'' or ``zone coverage'' (described in Section \ref{sec:background}) to the identified groups.  Our approach -- generating an information-rich feature set and using mixture models to assign labels -- can be adapted to almost any event or strategy annotation problem in the NFL, provided a set of tracking data.  For example, this approach could be used to automatically identify run-blocking strategies (e.g. guard pulls), pass rush assignments, pass rush strategies (e.g. stunts, blitzes, etc), and other sub-problems.

Finally, the use of mixture modeling (or ``model-based clustering'') comes with several benefits.  Most importantly, mixture models provide an actual statistical model to describe the cluster structure for a given problem.  Because it is a density-based clustering approach, there are no heuristics involved in the estimation of the model.  One advantage to density-based methods is that the resulting clusters are typically more interpretable, since we can characterize each cluster by the features of the density we fit to the data (e.g. each group's mean and variance in mixture modeling).  Additionally, mixture models provide ``soft'' cluster assignments -- i.e. probabilities of membership in each cluster -- allowing us to quantify the uncertainty in each cluster assignment.

The rest of this paper is organized as follows:  We first describe defensive pass coverage schemes for passing plays in football in Section \ref{sec:background}.  Then, we provide detail on each step of our clustering process in Section \ref{sec:methods}.  We detail the feature set that we design for the purposes of clustering pass coverage types in Section \ref{sec:features}.  We describe mixture modeling in Section \ref{sec:cluster}, highlighting the features of this technique that make it suitable for identifying pass coverage types.  We describe our approach for evaluating our clustering results and assessing the utility of each feature in Section \ref{sec:ari}.  We present the results of these approaches in Section \ref{sec:results}, and we discuss future directions for this work in Section \ref{sec:discussion}.

%Despite the clear importance of the defensive backs in the passing game, there exists no real meaningful annotations of what the cornerbacks and the safeties are doing in the secondary for pass coverage. The goal of this paper is to provide meaningful and interpretable annotations of what both of these position groups are doing in pass coverage in a purely unsupervised method. We attempt to solve this problem through feature generation of the player tracking data and clustering through Gaussian mixture models and hierarchical clustering techniques. The fundamental idea behind this approach is that features of a defensive players movement likely plays a major role in determining their defensive coverage scheme, and these features can be studied to make real time decisions on the field. 

\section{Characterizing Pass Coverage Types in Football}
\label{sec:background}

A defensive back in football is a player who lines up in the defensive backfield (typically at least a yard past the line of scrimmage, and to the outside of the field).  Their primary objective is to prevent the offense from completing any passes by covering wide receivers (offensive players who typically line up on the outside).  In this section, we distinguish the different types of defensive backs, and we describe the two primary types of individual pass coverage that defensive backs play.  

\subsection{Types of Defensive Backs}

There are two different, specialized positions broadly classified as defensive backs:  cornerback (CB) and safety (split into ``strong safeties'', SS, and ``free safeties'', FS).  

Generally, cornerbacks are more adept at providing close coverage on wide receivers and defending passes.%, and less adept at making tackles in the open field.  
The position requires speed and agility, and the ability to track a receiver (in man coverage) or occupy a space and read the quarterback (in zone coverage).  

\begin{figure}[t!]
  \centering
  \includegraphics[keepaspectratio, width=0.6\textwidth]{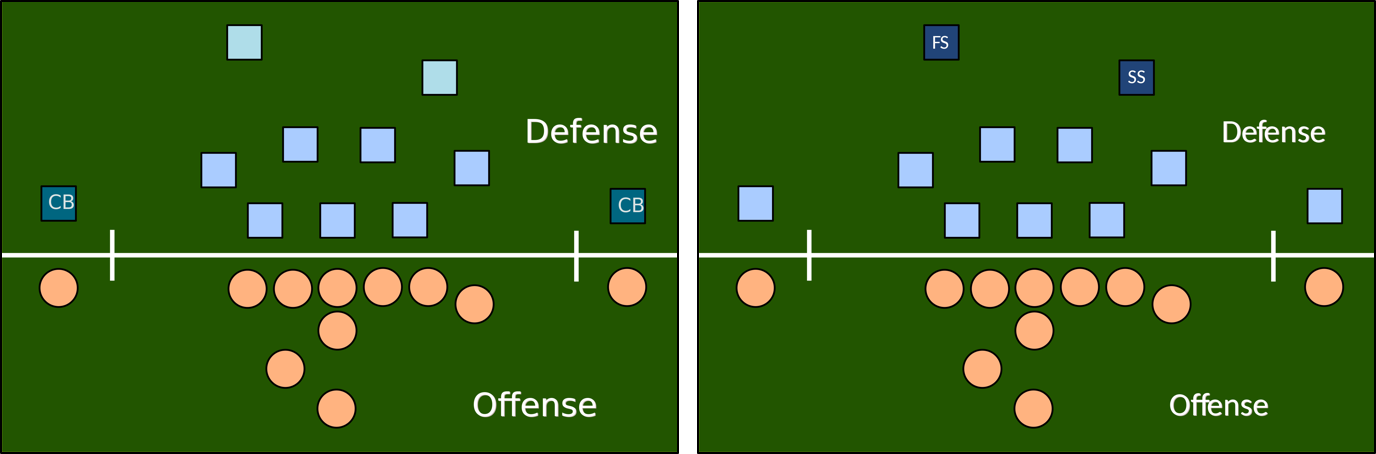}
  \caption{Defensive Backs in Football. The left-most and right-most blue squares are cornerbacks (labeled clearly on the left figure); the top-most two blue squares are safeties (labeled clearly in the right figure).  Cornerbacks and safeties are primariliy responsible for pass coverage.}
  \label{fig:defbacks}
\end{figure}

Safeties usually start the play 10-15 yards beyond the line of scrimmage and can be thought of as the last line of defense.  Their roles often depend on how the personnel is used by the offensive team:  sometimes, they provide additional help in defending long passes; other times, they provide man coverage on additional offensive receivers that are on the field.  They are also responsible for reading the play, quickly determining if it is a run or pass play, and reacting accordingly.  The typical alignment of players at each of these positions is shown in Figure \ref{fig:defbacks}.

In this paper, we focus primarily on pass coverage types of cornerbacks, which typically fit into one of two categories:  man coverage and zone coverage.  Below, we characterize each type of coverage and offer insight into the types of features that will be useful in distinguishing man coverage from zone coverage.

\subsection{Types of Individual Defensive Pass Coverage}

Individual defensive backs typically play one of two types of pass coverage:  man coverage and zone coverage, which we describe below.

In man coverage, a defensive back is assigned to defend a specific offensive player (typically a wide receiver).  Throughout the play, the defensive back follows that offensive player until the ball is thrown in an attempt to prevent that offensive player from making himself open and ultimately catching the ball.  In man coverage, the defensive back is focused on the movement of the offensive player, often with his head turned toward the offensive player, and often not even looking at the quarterback or the ball.  As such, the on-field motion of the defensive back may tend to mirror that of the offensive player that the cornerback is covering.  Importantly, this does not mean that the patterns of motion of defensive backs in man coverage will follow well-defined trajectories, as is the case for wide receivers.  Instead, their patterns of motion will correlate closely with those of the player they are covering. %A defender in man coverage attempts to read the receiver in order to determine when the ball is arriving rather than looking at the quarterback.
In the context of our problem, generating features that reflect this type of motion will be important.  This requires identifying the offensive player to which each defensive back is assigned to cover and building a set of features that capture this type of movement.

\begin{figure}[t!]
  \centering
  \includegraphics[keepaspectratio, width=0.6\textwidth]{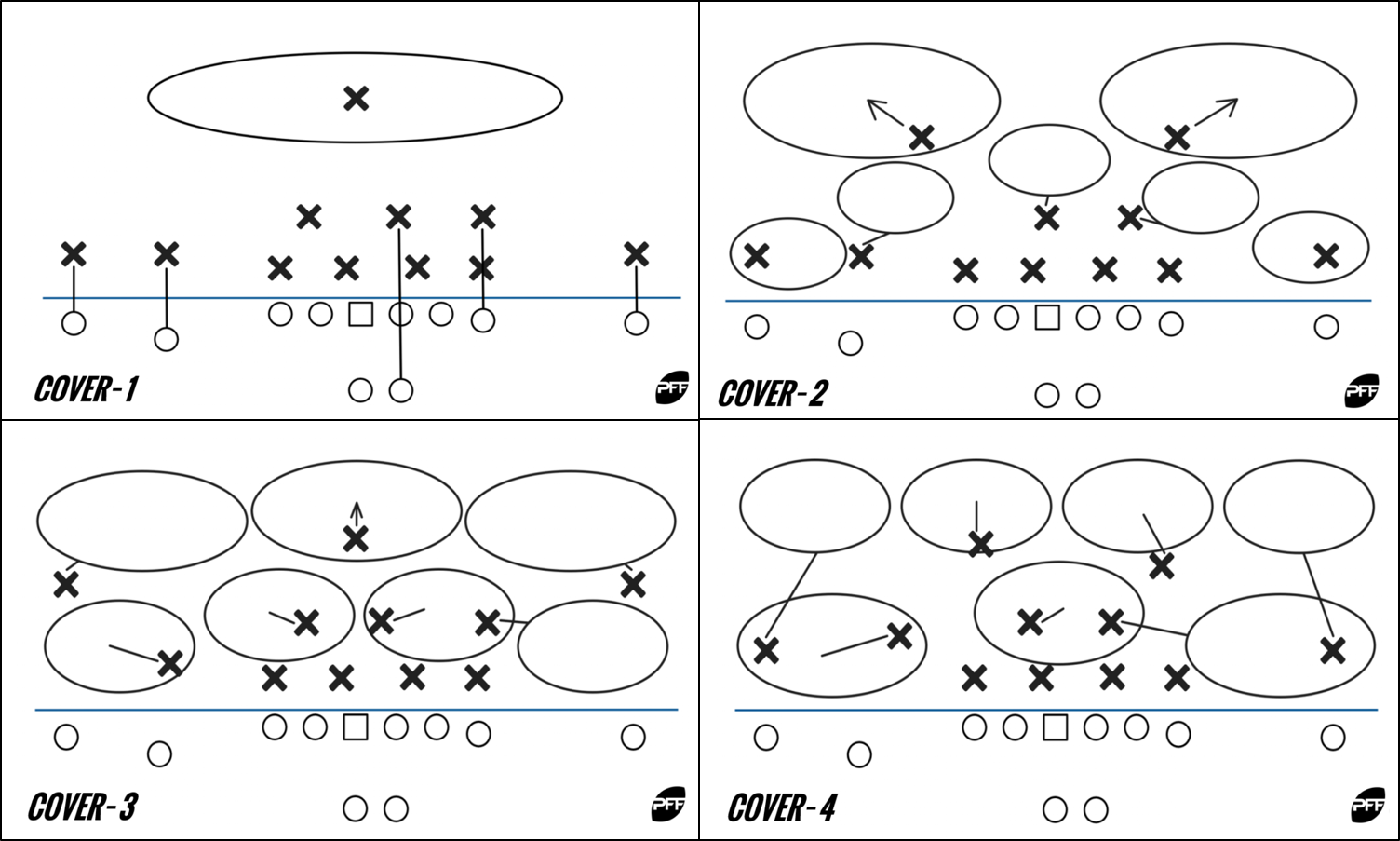}
  \caption{In different forms of Zone coverage, players are assigned different areas of the field to cover (denoted by the circle). In some types, certain players will be assigned man coverage as well (denoted by the line). ``Cover-$n$'' means that $n$ players (typically safeties) are playing deep zone coverage, while the remaining players may be assigned to either man coverage or zone coverage closer to the line of scrimmage.  Cover-4 may be used in situations when preventing deep passes is of utmost importance (e.g. 3rd-and-20, or during a two-minute drill).  Cover-1 my be used more frequently in short-yardage situations.  Cover-2 is probably the most common of these coverage schemes.}
  \label{fig:defbacks2}
\end{figure}

In zone coverage, a defensive back is assigned a zone on the field to defend.  These zones are constructed so that both the linebackers (who typically line up closer to the line of scrimmage and in the center of the field) and the defensive backs work together to provide a complete coverage of the possible passing areas in order to prevent the completion of a pass.  Generally, when a player is in zone coverage, their eyes are on the quarterback and their movements are less reactive to the movements of specific offensive players.  

This is not always the case, however.  For example, if a defensive back in zone coverage sees that there is only one offensive player near his zone, his responsibility shifts, so that he must now cover that single player in a way that resembles man coverage.  In other words, the locations and trajectories of teammates, opponents, and the ball can cause the defensive back to adjust his assignment \emph{in the middle of the play}.  This ``hybrid'' style of defensive coverage can be difficult to measure through feature generating, and can make the automatic identification of zone coverage challenging.  %The defenders are aware of the offensive players in their zone, but their focus is on reacting to the quarterback making a pass rather than focusing on the receiver. Many times, the movement of one defensive player indicates the type of coverage another player is in because of the nature of zone defense. 

Most of the time, teams employ some combination of man coverage and zone coverage on a single play.  For example, some teams would play a variation of Cover 2 (example shown in Figure \ref{fig:defbacks2}), where the cornerbacks are in man coverage and the safeties are in zone coverage.%This mixture of different coverage styles is done in order to attempt to confuse the offense or take advantage of certain personnel skill sets. Since the players are intentionally trying to fool the quarterback, their motion may represent this attempt, and categorizing their defense may be best done through a Gaussian Mixture Model.

\section{Methods}
\label{sec:methods}

In this section, we define a set of features describing the motion of each defensive back, and then apply several clustering methods to these features. In this way, we attempt to encapsulate the motion of the defensive backs and make a distinction between different types of coverage in an entirely unsupervised way.

\subsection{Data}

We use NFL player and ball tracking data from the NFL's inaugural Big Data Bowl.  This dataset consists of game data from the first six weeks of the 2017 NFL season. Each play from each game uses the league's player and ball tracking technology to record the locations and trajectories of all 22 players on the field (and the ball) throughout the duration of the play, at a rate of 10 Hz (10 frames per second). For each play, data is recorded starting from when the offense is set\footnote{Because of this, the pre-snap motions of offensive players are included, and the corresponding reactionary movements of defensive players are captured, allowing us to use this information when identifying coverage types before the snap (similar to how a quarterback will read a defense by putting receivers or running backs in motion before the snap).}.  For each player, each frame contains their $x$ and $y$ coordinates on the field with $0 \leq x \leq 120$ and $0 \leq y \leq 58$ for each frame. Furthermore, for each player, each frame consists of their speed $s$, displacement from last position, and direction of motion on the field described by the angle $\theta$, $0 \leq \theta \leq 360$.  A subset of frames are also labeled with text indicating the on-field event that happened at that frame (e.g. ball is snapped, first contact, etc).  For the purposes of this paper, the relevant events are mostly \texttt{ball\_snap} and \texttt{pass\_forward}.  %Metadata about individual game attributes is also available, however this paper will not use that data. 

One piece of information that is not available to us is the orientation of the defensive players on the field.  This is unfortunate, since the direction that a defensive back faces (relative to the offensive player to which he is assigned) would likely be one of the most informative features when distinguishing man vs. zone coverage.  We discuss this further in Section \ref{sec:discussion} and offer insight into how this additional information may be used by NFL teams, who have access to this data.

We use this data to generate features characterizing the movement of the defensive back, such that a clustering of these features will result in a meaningful interpretation of their coverage type.  %Furthermore, each player has an official position. This paper will work to identify the motion of the players that are labelled \texttt{CB}, \texttt{FS},or \texttt{SS}.  
There are 6,712 pass plays that we examine from the data set. From these 6,712 pass plays we generate 16,316 feature vectors in the feature generation step.

\subsection{Feature Generation}
\label{sec:features}

The relationships between the type of pass coverage (man or zone) and the movements, positions, and trajectories of players (relative to offense or otherwise) change at different time periods throughout the play.  For example, if a cornerback is facing the line of scrimmage at the start of a play, this tells us nothing about the type of coverage he is playing.  However, if he is still facing the line of scrimmage at the time the ball is thrown, he is more likely to be playing zone coverage.  For this reason, we design features that can be estimated at different points throughout the play that correspond to on-field events.  

\begin{figure}[]
  \centering
  \includegraphics[keepaspectratio, width=0.8\textwidth]{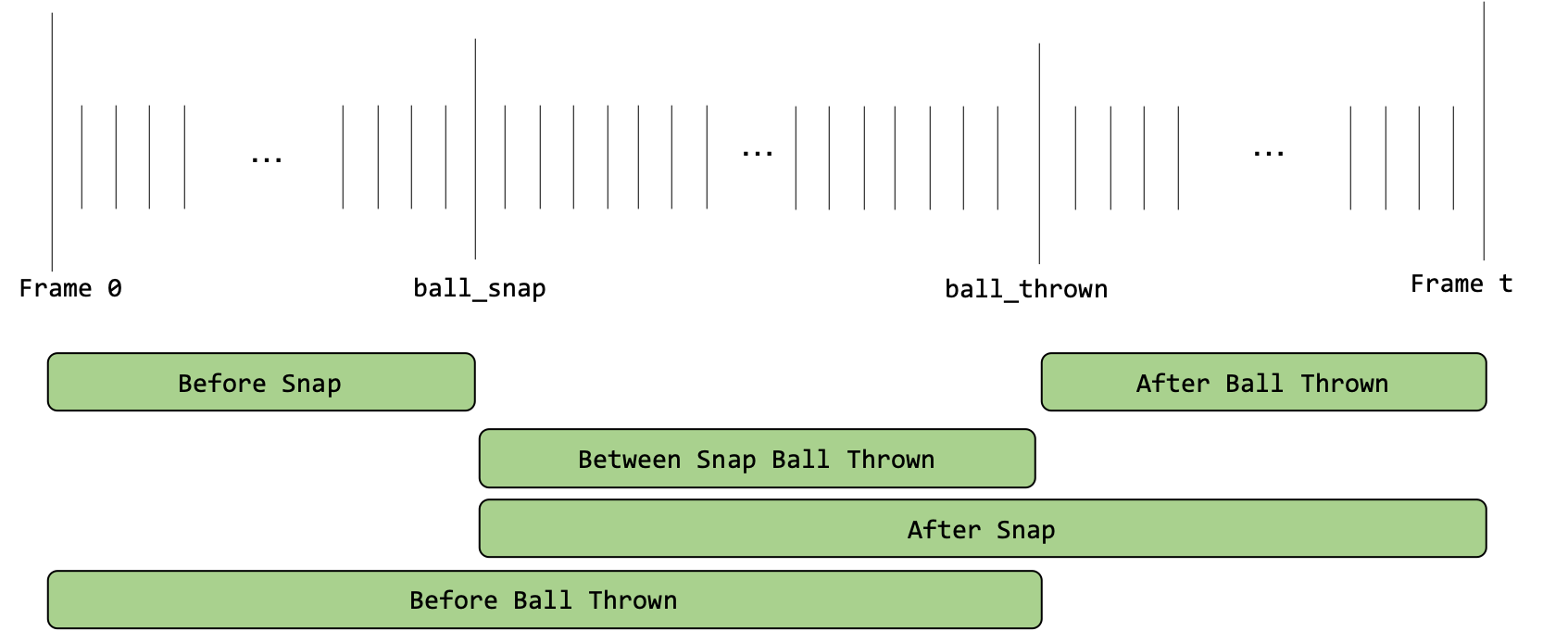}
  \caption{The time periods over which features are extracted}
  \label{fig:timesplit}
\end{figure}

%The main time of motion that is analyzed is exactly the time between the ball being snapped (\texttt{ball\_snap}) and the ball being thrown (\texttt{pass\_forward}). We choose to do this because the assumption is after the ball is thrown players are no longer in coverage, but rather attempting to defend the targeted receiver or pursue the player who catches the ball.  The movements of these players during this portion of the play should be independent of their coverage assignment, and thus may not provide as much value in the clustering model. However, over 4 other time periods as shown in figure \ref{fig:timesplit}, we generate the same 11 features. This will later allow us to model the evolution of probabilities over time. 

These time periods are described in Figure \ref{fig:timesplit}:  For each cornerback on each play, and for each of the time periods described in this figure, we generate a feature vector that consists of the features described in Table \ref{tab:features}.  The features are generated with respect to the cornerback being analyzed on each play.  Let $(x_{i,t}, y_{i,t})$ represent the coordinates of cornerback $i$, and $s_{i,t}$ represents his speed, all at frame $t$.  Let $O$ and $D$ refer to the set of offensive players and set of defensive players respectively. $m_{i,t}$ refers to player $i$'s direction of motion at frame $t$.  Let $T$ be the number of frames for the play in question (in practice, $T$ varies from play-to-play and would be indexed by the number of plays).  In the table below, $\mu_O$ ($\mu_D$) refers to the mean distance to the closest offensive (defensive) player, while $\mu_m$ refers to the mean difference in directions of motion, $\mu_x$ refers to the mean $x$-coordinate on a play, and so on.  Unless otherwise stated below, player $i$ refers to the cornerback in question on a given play (i.e. the defensive player for whom we are trying to predict the coverage type), player $j$ refers to the closest offensive player, and player $k$ refers to the closest teammate (another defensive player).

\begin{longtable}{| p{0.18\textwidth} | p{0.37\textwidth} | P{0.40\textwidth} |}
    \hline
    Feature Name & Feature Description & Feature Equation
     \\ \hline
    \texttt{VAR\_X} & Variance in the x coordinate & $\frac{\sum_{t=1}^{T}(x_{i,t} - \mu_x)^2} {n}$
    \\ \hline
    \texttt{VAR\_Y} & Variance in the y coordinate & $\frac{\sum_{t=1}^{T}(y_{i,t} - \mu_y)^2} {n}$
    \\ \hline
    \texttt{SPEED\_VAR} & Variance in the speed & $\frac{\sum_{t=1}^{T}(s_{i,t} - \mu_s)^2} {n}$
    \\ \hline
    \texttt{OFF\_VAR} & Variance in the distance from the nearest offensive player at every frame & $\frac{\sum_{t=1}^{T}(d_t - \mu_O)^2} {T}$, \scalebox{0.6}{$d_t = \argmin\limits_{j \in O}\sqrt{(x_{j,t} - x_{i,t})^2 + (y_{j,t} - y_{i,t})^2}$}
    \\ \hline
    \texttt{DEF\_VAR} & Variance in the distance from the nearest defensive player at every frame & $\frac{\sum_{t=1}^{n}(d_t - \mu_D)^2} {T}$, \scalebox{0.6}{$d_t = \argmin\limits_{k \in D}\sqrt{(x_{k,t} - x_{i,t})^2 + (y_{k,t} - y_{i,t})^2}$}
    \\ \hline
    \texttt{OFF\_MEAN} & Mean distance from the nearest offensive player at every frame & $\frac{\sum_{t=1}^{n}d_t} {T}$, \scalebox{0.6}{$d_t = \argmin\limits_{j \in O}\sqrt{(x_{j,t} - x_{i,t})^2 + (y_{j,t} - y_{i,t})^2}$}
    \\ \hline
    \texttt{DEF\_MEAN} & Mean distance from the nearest defensive player at every frame & $\frac{\sum_{t=1}^{n}d_t} {T}$, \scalebox{0.6}{$d_t = \argmin\limits_{k \in D}\sqrt{(x_{k,t} - x_{i,t})^2 + (y_{k,t} - y_{i,t})^2}$}
    \\ \hline
    \texttt{OFF\_DIR\_VAR} & Variance in the difference in degrees of the direction of motion between the player and the nearest offensive player & 
    $\frac{\sum_{t=1}^{T}((m_{j,t}-m_{i,t}) - \mu_m)^2} {T}$
    \scalebox{0.6}{$j = \argmin\limits_{j \in D}\sqrt{(x_{j,t} - x_{i,t})^2 + (y_{j,t} - y_{i,t})^2}$}
    \\ \hline
    \texttt{OFF\_DIR\_MEAN} & Mean difference in degrees of the direction of motion between the player and the nearest offensive player & 
    $\frac{\sum_{t=1}^{T}(m_{j,t}-m_{i,t})} {T} $ \scalebox{0.6}{$j = \argmin\limits_{j \in D}\sqrt{(x_{j,t} - x_{i,t})^2 + (y_{j,t} - y_{i,t})^2}$}
    \\ \hline
    \texttt{RAT-MEAN} ($\mu_r$) & Mean ratio of the distance to the nearest offensive player $j$ and the distance from the nearest offensive player to the nearest defensive player $k$ &
    \vspace{-20mm}
    \makecell{$\mu_r = \frac{\sum_{t=1}^{T}\frac{\sqrt{(x_{j,t} - x_{i,t})^2 + (y_{j,t} - y_{i,t})^2}}{\sqrt{(x_{j,t} - x_{k,t})^2 + (y_{j,t} - y_{k,t})^2}}} {T}$
    \\
    \scalebox{0.6}{$j = \argmin\limits_{j \in O}\sqrt{(x_{j,t} - x_{i,t})^2 + (y_{j,t} - y_{i,t})^2}$} \\ \scalebox{0.6}{$k = \argmin\limits_{k \in D}\sqrt{(x_{k,t} - x_{i,t})^2 + (y_{k,t} - y_{i,t})^2}$}}
    \\ \hline
    \texttt{RAT-VAR} & Variance of the ratio of the distance to the nearest offensive player and the distance from the nearest offensive player to the nearest defensive player $k$ & \vspace{-20mm}
    \makecell{$\frac{\sum_{t=1}^{T}(\frac{\sqrt{(x_{j,t} - x_{i,t})^2 + (y_{j,t} - y_{i,t})^2}}{\sqrt{(x_{j,t} - x_{k,t})^2 + (y_{j,t} - y_{k,t})^2}} - \mu_r)^2}{T}$,\\
    \scalebox{0.6}{$j = \argmin\limits_{j \in O}\sqrt{(x_{j,t} - x_{i,t})^2 + (y_{j,t} - y_{i,t})^2}$} \\ \scalebox{0.6}{$k = \argmin\limits_{k \in D}\sqrt{(x_{k,t} - x_{i,t})^2 + (y_{k,t} - y_{i,t})^2}$}}
    \\ \hline
    \caption{Features describing the movement of defensive backs in pass coverage.}
    \label{tab:features}
\end{longtable}

% We then take the created features and make a vector:
% $$\langle \texttt{VAR\_X},\texttt{VAR\_Y},\texttt{SPEED\_VAR},\texttt{OFF\_VAR},\texttt{DEF\_VAR},\texttt{OFF\_MEAN},\texttt{DEF\_MEAN},\texttt{OFF\_DIR\_VAR},$$ $$ \texttt{OFF\_DIR\_MEAN},\texttt{0\_DIS\_OFF},\texttt{2\_DIS\_OFF} ,\texttt{4\_DIS\_OFF},\texttt{RAT-MEAN},\texttt{RAT-MID},\texttt{RAT-PASS},\texttt{RAT-SNAP},\texttt{RAT-VAR}                  \rangle$$

From the features we generate, we would expect the two ratio (\texttt{RAT-X}) features to be most helpful in determining man vs. zone coverage, because in man coverage we would expect the defensive back to follow the offensive player he is covering very closely, whereas in zone coverage this is not necessarily true.  Since there is generally a ``hard'' assignment for each defender in man coverage (i.e. each defender has a specific player to cover), the ratio of the distance from the cornerback to the closest offensive player, to the distance from cornerback's closest teammate to the same offensive player should be fairly small. That is, we expect the cornerback in man coverage to have a very small distance to his assignment, and any of the other remaining defenders to have a comparably larger distance to that player. We compute this value throughout the course of the play, and summarize those values with five quantities: the mean and variance of this value, in order to summarize the changes in this value throughout the play; the value of this quantity at the snap; at the time the ball is thrown; and at the mid-point between these two timepoints. 

Furthermore, another feature we expect to differentiate coverage types is the \texttt{OFF\_DIR\_VAR} and \texttt{OFF\_DIR\_MEAN}. This is because, similar to the above logic, a player in man coverage would be following their assignment around the field and would be much more ``reactive'' to the motion of the offensive player. Hence we would expect the direction of motion to be almost the same as their assignment throughout the course of the play.  %\footnote{A lagged version of this may also prove valuable, as it may account for delays in reaction time.} 
For a player in zone coverage, they are generally watching the quarterback rather than watching a receiver, and thus their motion might be more static, so that the difference in the direction of motion to the nearest offensive player would both be different and would have more variance. 

\subsection{Mixture Models}
\label{sec:cluster}

Clustering is the process of partitioning observations in a dataset into groups without respect to some response variable \citep{Hartigan75}.  Often, this process is referred to as ``unsupervised learning,'' since clustering models can be used to assign labels for discrete groups without training data.  

Mixture modeling, or model-based clustering, is a type of clustering algorithm that fits a mixture of probability density functions to a dataset, where each density is representative of a single group or cluster.  A mixture model can be written in the form:

\begin{equation*}
  f(x) = \sum_{g = 1}^G \pi_g f_g(x | \rho_g)
\end{equation*}

\noindent where $g$ indexes over the groups (or clusters) of the mixture model), $f_g(x | \rho_g)$ represents the density for group $g$ with parameters $\rho_g$, and $f(x)$ is the overall mixture distribution.  

Commonly, Gaussian mixtures (with $f_g$ representing a Gaussian probability density function) are used because of their ease of implementation and theoretical properties.  That said, any parametric distribution (or mixture of parametric distributions) can be used \citep{mbc}.  The choice of $G$ is left to the user, and is commonly determined by searching over a range of possible values and determining the best with an evaluative measure like the Bayesian Information Criterion (BIC), though many methods for doing this exist.  We direct interested readers to \cite{McNicholas2016} for a complete overview of model-based clustering and the related literature.

In this paper, we use Gaussian mixture modeling with the features described in Section \ref{sec:features} to provide a model distinguishing man coverage from zone coverage.  We take $f_g$ to be Gaussian densities, and we test several different values of $G$, with results provided in Section \ref{sec:results}.  For the remainder of this paper, we refer to our Gaussian mixture modeling approach as GMM.  

There are two key benefits to using mixture models, for our purposes.  First, mixture models yield ``soft'' cluster assignments (i.e. probabilistic labels for each cluster), allowing us to quantify how certain we are when assigning man coverage or zone coverage labels.  Second, mixture models are density-based, statistical models that estimate a empirical probability distribution from real data.  As such, they come with several helpful properties.  First, this allows us to characterize the resulting clusters in informative ways.  For example, we can describe the $k$-dimensional centroid of each cluster, or the values of each feature that correspond to the center of each cluster.  In the context of our problem, this allows us to describe how each of our features distinguish man vs. zone coverage.  Second, we can obtain cluster membership probabilities from the mixture model for \emph{any} observation, including data from a holdout set of data.  In the context of football, this means that we can use our mixture model to make man vs. zone predictions for each defensive back on each play in real-time throughout the course of a football game, even though this game as not included in the data on which we estimated the model.  Third, the probabilistic cluster labels allow us to provide an extensive evaluation of the clustering results, as we demonstrate later.

\subsection{Evaluation of Clustering Results with Cross-Validation}
\label{sec:ari}

Many approaches exist for evaluating clustering results, but there is no ``ground truth'' data to which we can compare our clustering results.  Moreover, even if an extensive set of labels existed, \cite{Hennig} points out that ``The fact that we know certain true classes doesn't preclude other legitimate, or `true' clusterings ... There could be better truths than the known one.'' In other words, Hennig argues that clustering results should not always be evaluated against a set of ground truth labels, but by a critical evaluation of the model itself, and the features used to estimate the model.  We take this approach below for choosing the number of clusters $G$ (Section \ref{sec:choose-g}) and assessing the influence of features (Section \ref{sec:feature-analysis}).

We utilize the fact GMMs are fitting a statistical model for which we can generate (i.e. predict) cluster labels for data points to evaluate our clustering results. We use ``leave-one-week-out'' cross-validation (LOWO CV) to compare test set partitions from mixture models fit on training data versus test data. For each week $k \in \{1,\hdots, K\}$), we proceed as follows:
\begin{enumerate}
    \item fit GMM on training data $\hat{f}^{\text{train}}$ using weeks $\{1,\hdots, K\} \backslash \{k\}$, 
    \item fit GMM on test data $\hat{f}^{\text{test}}$ using week $\{k\}$,
    \item create partitions $p_k^{train}$ and $p_k^{test}$, using $\hat{f}^{\text{train}}$ and $\hat{f}^{\text{test}}$ respectively, for each observation in holdout week $k$ by assigning cluster labels:
    \begin{itemize}
        \item let $\hat{f}^*$ denote an arbitary GMM (either $\hat{f}^{\text{train}}$ or $\hat{f}^{\text{test}}$), 
        \item for a given point $x \in \mathbb{R}^d$, obtain the probability density from the estimated mixture density $\hat{f}^*_g$ for each label $g$,
        \item compute $P(G(x) = g) = \frac{\hat{f}^*_g(x)}{\sum_g \hat{f}^*_g(x)}$, where $G(x)$ indicates the label of $x$,
        \item assign cluster label for $x$ with $g^* = \argmax_g P(G(x) = g)$.
    \end{itemize}
    \item compare GMMs as a function of their holdout week partitions  $h_k(p_k^{train}, p_k^{test})$.
\end{enumerate}

For our purposes, we consider $h$ to be a function of pair agreements between the respective holdout partitions (i.e. for a given GMM $\hat{f}^*$ do points $x_i$ and $x_j$ have same or different cluster label?). Following \citet{Steinley04}, our notation for the pair agreements between our holdout partitions of interest $p_k^{train}$ and $p_k^{test}$ is presented in Table \ref{tab:ari}. A simple comparison is to use the Rand index \citep[RI]{Rand71}, which is just the ratio of pair agreements to the total number of pairs, $\text{RI}_k(p_k^{train}, p_k^{test}) = \frac{A + D}{N}$, where $N = A + B + C + D$. However, since the RI value can be inflated due to chance agreement we use the adjusted Rand Index \citep[ARI]{ari} as our function $h$ compare holdout partitions,
\begin{equation*}
    \text{ARI}_k(p_k^{train}, p_k^{test}) = \frac{N(A + D) - [(A + B)(A + C) + (C + D)(B + D)]}{N^2 - [(A + B)(A + C) + (C + D)(B + D)]}.
\end{equation*}
Under completely random partitions $\mathbb{E}[\text{ARI}_k(p_k^{train}, p_k^{test})] = 0$. This establishes an appropriate baseline for evaluating our clustering results, with $ARI(p_k^{train}, p_k^{test}) < 0$ indicating that our GMMs fit separately on training versus test data are leading to worse than random cluster labels. Observing the maximum value of 1 indicates that the two holdout partitions being compared are identical to each other. Thus, we seek to fit GMMs that maximize $\text{ARI}_k(p_k^{train}, p_k^{test})$ indicating a detection of clustering structure that is in agreement in both the training and test weeks of our LOWO CV procedure. While we compare two separate GMMs fit on the training and test data, one could replace either GMM above with any labeling process (e.g. expert labels of coverage). We direct readers to \citet{Steinley04} for an assessment of the properties of ARI, such as its advantage over the use of misclassification rate when the number of cluster labels equals the number of ``true'' labels.

\begin{table}[ht]
\caption{Notation for the cross-tabulation of observation pairs from holdout partitions $p_k^{train}$ (rows) and $p_k^{test}$ (columns).}
\label{tab:ari}  
\centering

\begin{tabular}{l|c|c}
  \hline
 & $p_k^{test}$ - same label & $p_k^{test}$ - different label  \\\hline
$p_k^{train}$ - same label & A & B \\ \hline
$p_k^{train}$ - different label & C & D \\
   \hline
\end{tabular}
\end{table}

\subsection{Determining Number of Clusters}
\label{sec:choose-g}

We choose the number of clusters $G$ using the LOWO CV procedure presented in Section \ref{sec:ari}. For each $G = 2,\hdots, 9$, we apply the LOWO CV procedure, resulting in $K = 6$ different ARI values, and compute the average across the holdout weeks:
\begin{equation*}
    \overline{\text{ARI}}_{G} = \sum_{k}^{K} \text{ARI}_{G,k}(p_{G,k}^{train}, p_{G,k}^{test}).
\end{equation*}
We then pick the number of clusters yielding the highest average LOWO CV ARI across all weeks, $G^* = \underset{G}{\arg\max}\ \overline{\text{ARI}}_{G}$.

\subsection{Assessment of Feature Influence}
\label{sec:feature-analysis}

Understanding the influence of different features in clustering and mixture models remains an open problem. A popular approach for GMMs casts the variable selection problem as a model-selection problem, greedily choosing which features to include based on BIC \citep{Raftery06} (see \citet{Fop2018} for an extensive review of variable selection for model-based clustering). Due to the number of features considered in Table \ref{tab:features}, rather than implement a greedy search, we use our LOWO CV framework from Section \ref{sec:ari} to measure each feature's influence on the clustering results as follows for a fixed number of clusters $G$: 
\begin{enumerate}
    \item Compute average LOWO CV ARI using all features, $\overline{\text{ARI}}^{all}_{G^*}$,
    \item for each feature $m \in \{1, \hdots M\}$:
    \begin{enumerate}
        \item compute average LOWO CV ARI with feature $m$ removed, $\overline{\text{ARI}}^{(-m)}_{G^*}$,
        \item compute difference in average LOWO CV ARI with and without feature $m$ as a measure of its clustering influence:
        \begin{equation*}
            \textit{Influence}_m = \overline{\text{ARI}}^{all}_{G^*} - \overline{\text{ARI}}^{(-m)}_{G^*}.
        \end{equation*}
    \end{enumerate}
\end{enumerate}
We then rank each feature by its LOWO CV $\textit{Influence}_m$ in Section \ref{sec:feature-eval}, similar to feature importance plots commonly used for various supervised learning models. We interpret more positive values for $\textit{Influence}_m$ as indicating that the inclusion of feature $m$ leads to an improvement in the quality of the clustering results

\section{Results}
\label{sec:results}

We first assess the number of clusters in Section \ref{sec:n-clusters} and then the influence of each feature in improving the clustering results in Section \ref{sec:feature-eval}, using the LOWO CV procedures explained in Sections \ref{sec:ari}-\ref{sec:feature-analysis}. Next, we demonstrate the practical meaning of our approach and interpretation of results in the context of football. We closely examine several specific plays in Section \ref{sec:plays}, demonstrating how the predicted coverage type probabilities can change throughout the course of a play. We draw connections to common football strategies, and we explain why mixture models are ideal for this type of analysis. Finally, we provide an analysis of coverage types by player, team, and game situation (Section \ref{sec:cluster-eval}).

For all results in this section, we apply the clustering models to only cornerbacks, since the features distinguishing the type of coverage by safeties should differ from those of cornerbacks.  That said, the same process could be applied independently to safeties. All clustering results were generated using the \texttt{GaussianMixture} function in the Python \texttt{scikit-learn} library \citep{scikit-learn}, which uses the expectation-maximization (EM) algorithm to estimate the mixture model parameters.  Due to the high degree of colinearity in our feature space, we use an unconstrained ``VVV'' covariance structure when fitting the models \citep{mbc}.   %This exercise is left to future work.

For the results in Sections \ref{sec:n-clusters}, \ref{sec:feature-eval}, and \ref{sec:cluster-eval}, we fit a mixture model using all features defined in Table \ref{tab:features} for all time periods defined in Figure \ref{fig:timesplit} (i.e. each feature replicated over each of the five time periods).  For the results in Sections \ref{sec:plays} and \ref{sec:results-analysis}, we fit mixture models separately for each time period, using only use the features available at the relevant time period of the play.

\subsection{Number of Clusters $G^*$}
\label{sec:n-clusters}

\begin{figure}[h]
  \centering
  \includegraphics[keepaspectratio, width=0.8\textwidth]{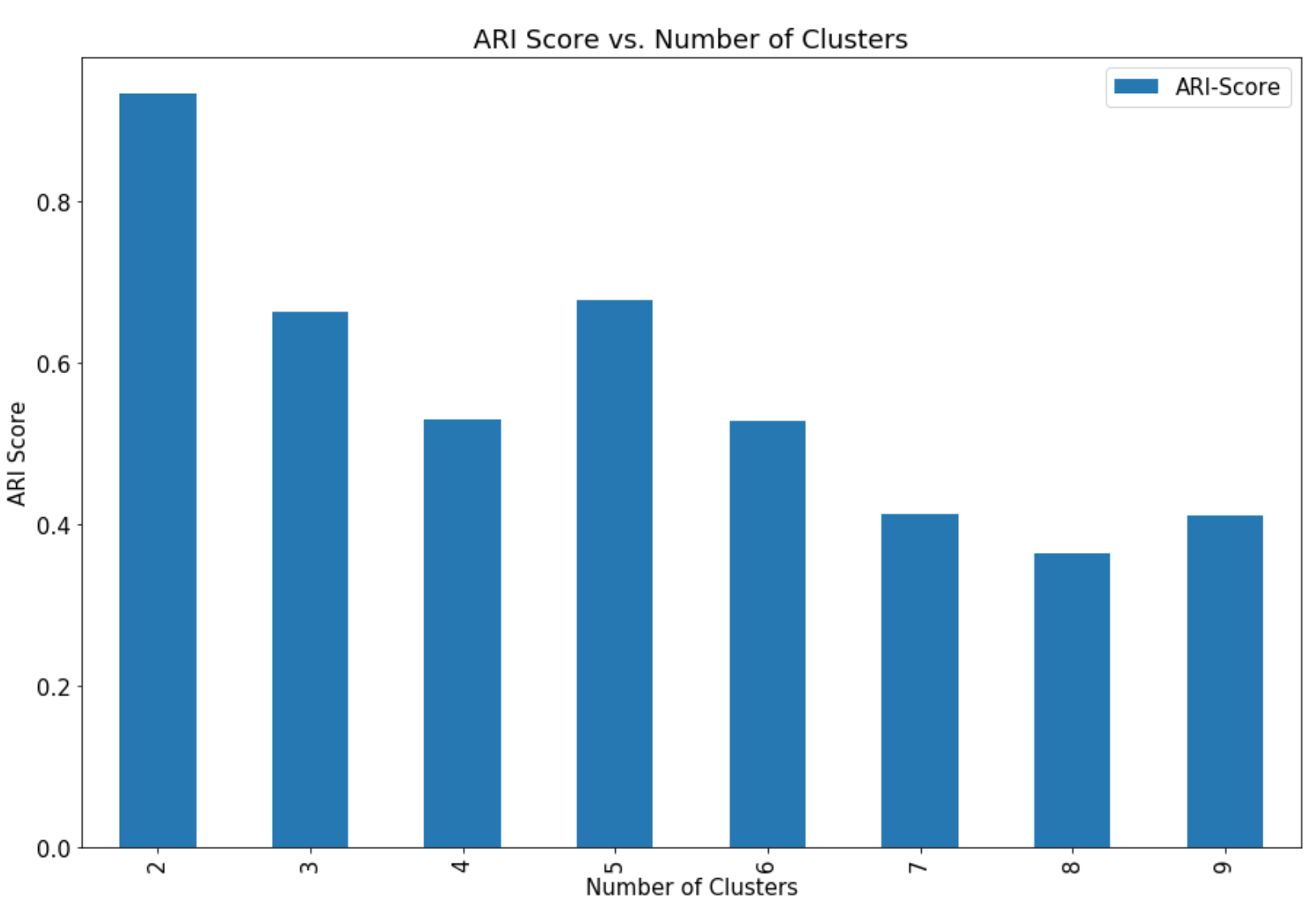} 
  \caption{The ARI scores for the number of clusters from our features}
  \label{fig:ari}
\end{figure}

First, we determine the number of clusters $G^*$ using the LOWO CV average ARI across the holdout weeks as explained in Section \ref{sec:choose-g}. As displayed in Figure \ref{fig:ari}, we found that $G^* = 2$ yields the highest value by a fairly wide margin, with an ARI over 0.9. This indicates that a two-cluster solution is the best fit for our data given our definition for evaluation in Section \ref{sec:ari}, which matches intuition in that there are primarily two types of cluster labels representing man and zone coverage.

Because the mixture model's clusters are completely data-driven, they do not come with common-sense labels like ``man coverage'' or ``zone coverage''.  Instead, we must make some determination of the type of coverage visually.  We manually examined many animations of plays and determined whether each cornerback on each play was playing man, zone, or unknown coverage\footnote{In an appeal to our own authority, we note that the co-author who undertook this task has experience playing defensive back in high school football.}.  Using this information in conjunction with the cluster membership probabilities, it is fairly easy to tell which cluster corresponds to man coverage and which corresponds to zone coverage.  We provide illustrations of how the coverage types can be visually distinguished throughout the course of a play in Section \ref{sec:plays}.

\subsection{Feature Influence on Clustering Results}
\label{sec:feature-eval}

\begin{figure}[h]
  \centering
  \includegraphics[keepaspectratio, width=0.7\textwidth]{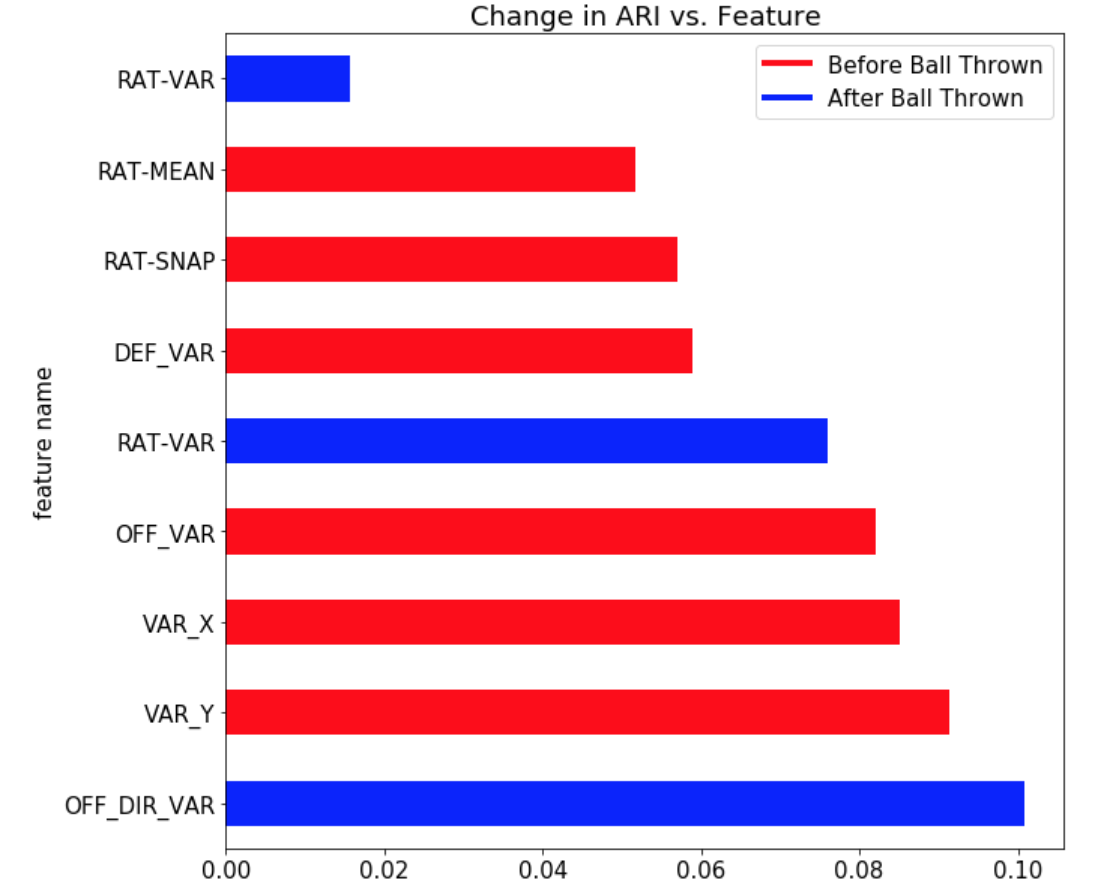} 
  \caption{Top nine features by LOWO CV influence on clustering results (from bottom to top).}
  \label{fig:ari-var}
\end{figure}

We proceed to measure $\textit{Influence}_m$ for each of our considered features in Table \ref{tab:features} using the LOWO CV procedure explained Section \ref{sec:feature-analysis}. Figure \ref{fig:ari-var} displays the top nine features in terms of the LOWO CV $\textit{Influence}_m$.  The majority of the top variables are measured before the ball is thrown, while the top feature is the variance in the difference of direction of motion between the player and the nearest offensive player after the ball is thrown. 

%Figure \ref{fig:ari} shows the results of this procedure.  The two-cluster solution for GMM yields by far the best results in this predictive setting, with a very high ARI of over 0.9.  This indicates that the GMMs fit on random train/test splits are in very strong agreement when the training set GMM is used to predict the cluster labels from the out-of-sample test set.  For the remainder of this paper, we focus on the two-cluster solution.

\subsection{Examining Soft Cluster Membership Probabilities}
\label{sec:cluster-eval}

\begin{figure}[h]
  \centering
  \includegraphics[keepaspectratio, width=0.4\textwidth]{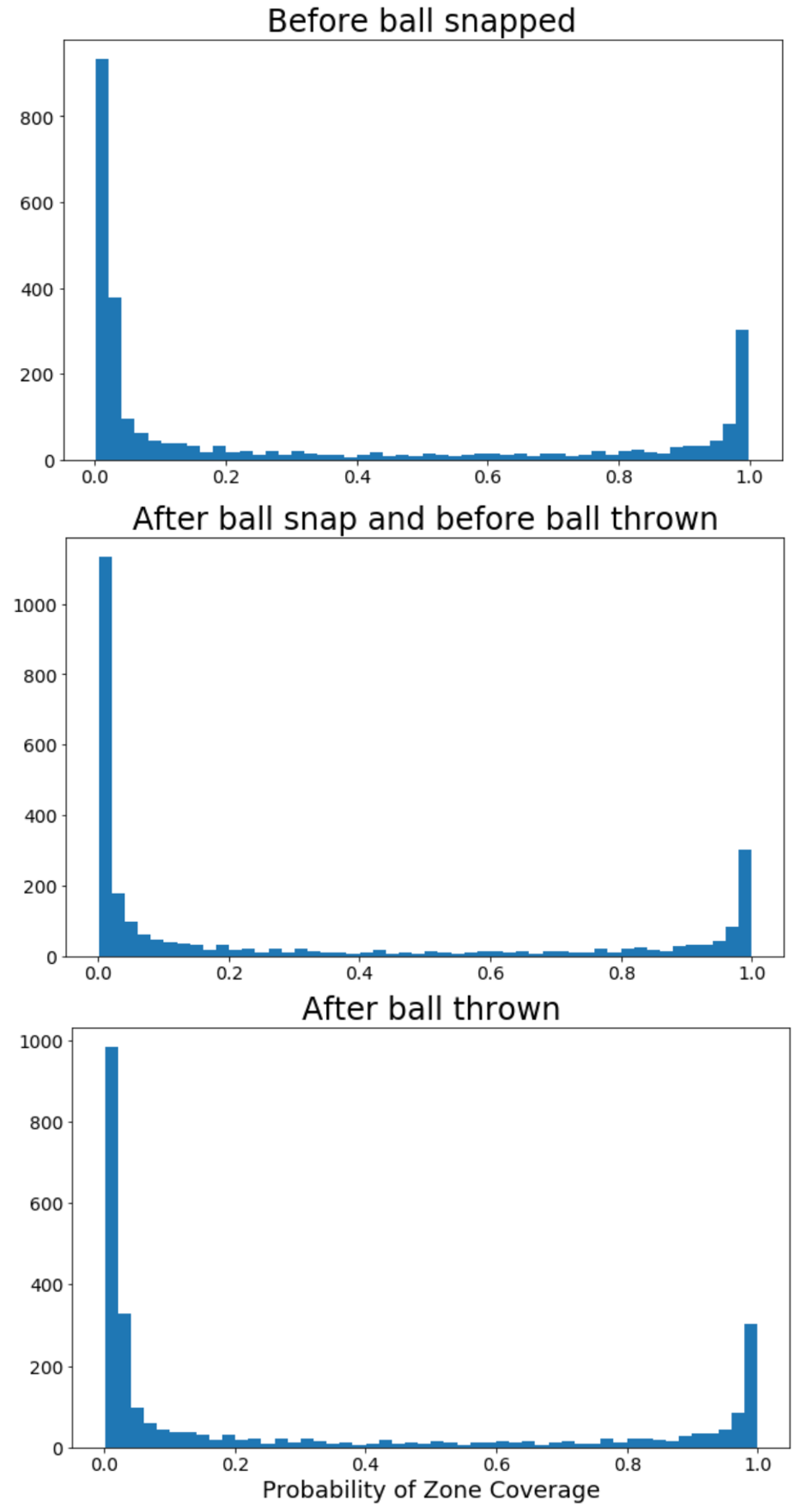}
  \caption{Distribution of cluster membership probabilities for zone coverage cluster.}
  \label{fig:distribution}
\end{figure}

Figure \ref{fig:distribution} shows the distribution of predicted probabilities of membership in the ``zone coverage'' cluster from the mixture model.  (The corresponding chart for man coverage, of course, is simply the reverse of Figure \ref{fig:distribution}.)

Here, we see similar distributions of cluster membership probabilities at different points throughout the play.  Man coverage is more common the zone coverage, and the mixture model typically yields cluster membership probabilities near the extremes, indicating that the two coverage type clusters are well-separated.  Before the ball is snapped, man coverage predictions are slightly more common than at other points throughout the play.  As discussed in Section \ref{sec:plays}, this may relate to how teams disguise their coverage at the beginning of plays, and so there is less variation in the alignment of defenders before the ball is snapped.

\subsection{Characterizing Pass Coverage Throughout Plays}
\label{sec:plays}

In modern defensive schemes, teams may try to disguise their pass coverage at the start of the play, in order to confuse the opposing quarterback.  Other schemes can involve a ``read and react'' approach to pass coverage, where specific coverage assignments depend on what happens at the start of the play.

Interestingly, the mixture model does a good job of recognizing these scenarios.  Coverage type probabilities often change substantially from the time of the snap to when the play has had a few seconds of ``burn-in'' time.  Some examples are given below. In each of these examples, the offense is in red, and the defense is in blue. The arrows point to the players we focus on. 

\begin{figure}[]
  \centering
  \includegraphics[keepaspectratio, width=0.8\textwidth]{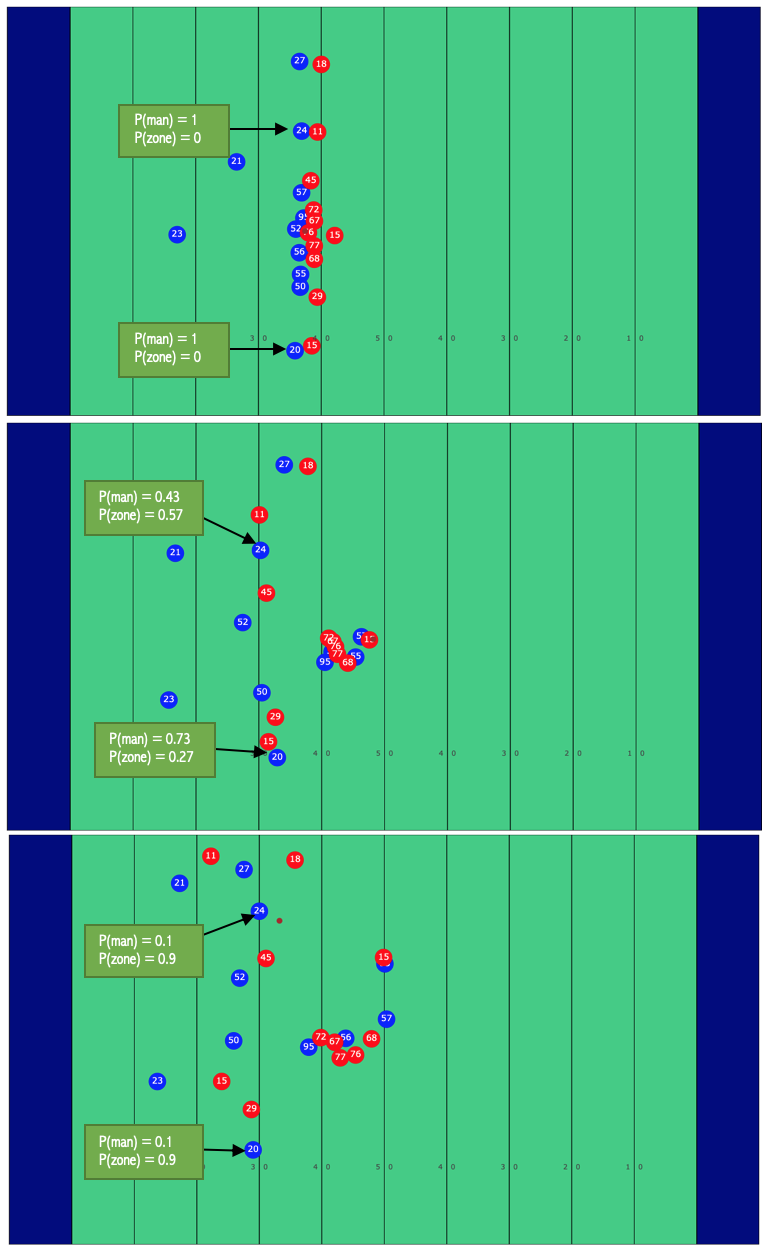} 
  \caption{Play 3765 from Game ID 2017091000, with offense shown in red, and defense shown in blue.  The play starts with the formation in the first frame and the ball is thrown in the last frame. The middle frame is exactly in between. In this case the top arrow points to player with jersey 24 (Defense-24) and the bottom to player with jersey 20 (Defense-20).}
  \label{fig:play1}
\end{figure}

In Figure \ref{fig:play1}, we can see the probabilities assigned to man and zone coverage at each frame throughout the play by the GMM. By the end of the play, the GMM predicts that Defense-24 (top arrow) and Defense-20 (bottom arrow) to have been in zone coverage.  We can see that they actually start out lined up as if they were going to play ``press'' coverage (a type of man coverage where the defensive back physically prevents the receiver from beginning his route at the start of the play), leading the GMM to predict both players to be playing man coverage with high probability.  As the play develops, Defense-20 and Defense-24 do not follow the men they initially lined up against and instead occupy their assigned zones.  We see this change in the second and third frames, when they clearly are in zone coverage (as they are occupying a space rather than following another offensive player).  The model reflects this with high probability.  In this play, the defense was likely disguising their coverage type at the start of the play in an attempt to trick the quarterback into misreading the defense.  Our mixture modeling approach quantifies this probabilistically as the play develops, until it is obvious that the defensive backs are actually playing zone coverage.

\begin{figure}[]
  \centering
  \includegraphics[keepaspectratio, width=0.8\textwidth]{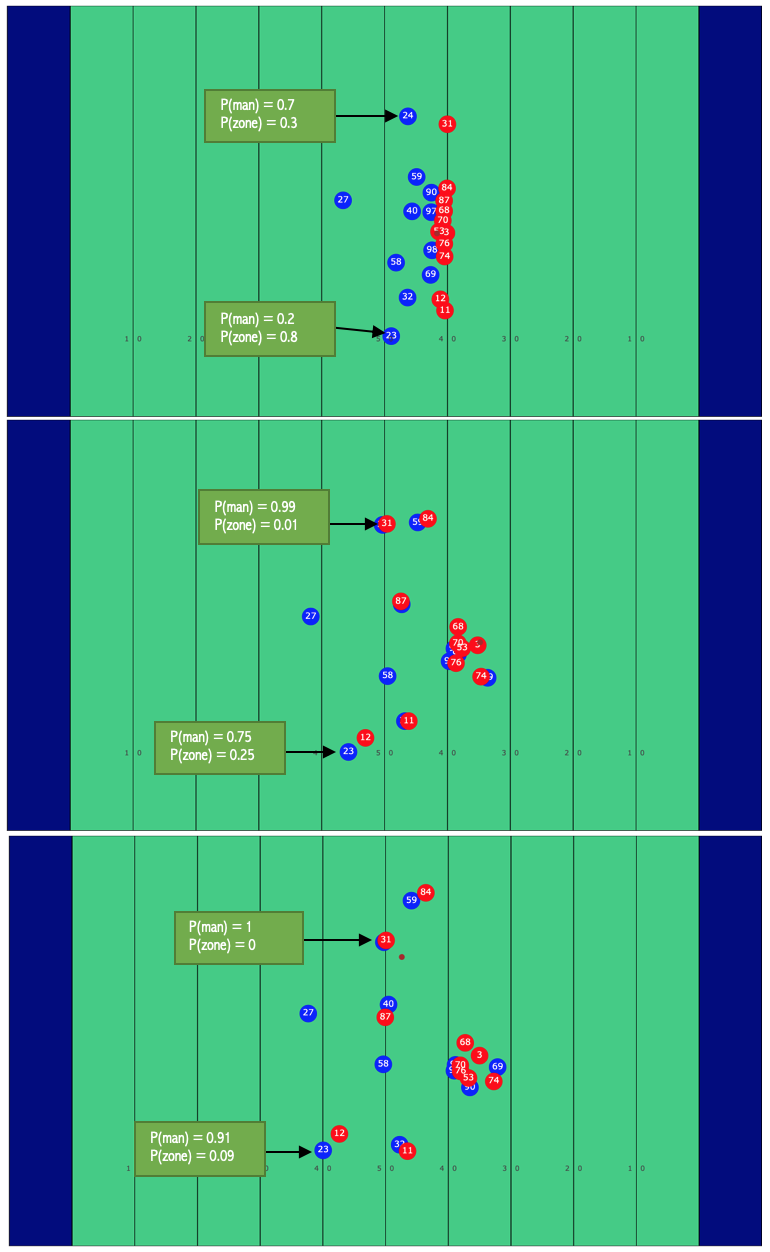}
  \caption{Play 174 from Game ID 2017091004, with offense shown in red, and defense shown in blue.  The play starts with the formation in the first frame and the ball is thrown in the last frame. The middle frame is exactly in between. In this case the top arrow points to player with jersey 24 (Defense-24) and the bottom arrow points to player with jersey 23 (Defense-23).}
  \label{fig:play2}
\end{figure}

In Figure \ref{fig:play2}, the offense is lined up with three receivers, and the defense does not line up in a way that indicates man coverage with high probability:  Each of the players that we are tracking are lined up slightly farther away from the player they would be assigned to cover, possibly indicating that they are playing zone coverage.  The GMM reflects this, cautiously predicting zone for Defense-23 ($p$ = 0.8) and man for Defense-24 ($p$ = 0.7).  Before the ball is snapped, Defense-23 is lined up relatively far away from a corresponding offensive player, leading to the model's prediction of zone coverage.  Similar to the previously discussed play, this was likely an attempt by the defense to disguise the coverage scheme being used.  As the play develops, however, we see that both defensive backs follow a specific offensive player until the ball is thrown, indicating that they were actually playing man coverage the whole time.  This change in coverage type probability from thee GMM is reflected in the middle frame, as the probability of man coverage sharply rises for both players, and finishes in the final frame with high probabilities of man coverage for both players.  %Thus the model was able to pick up clear instances of man coverage as well. 

\begin{figure}[]
  \centering
  \includegraphics[keepaspectratio, width=0.8\textwidth]{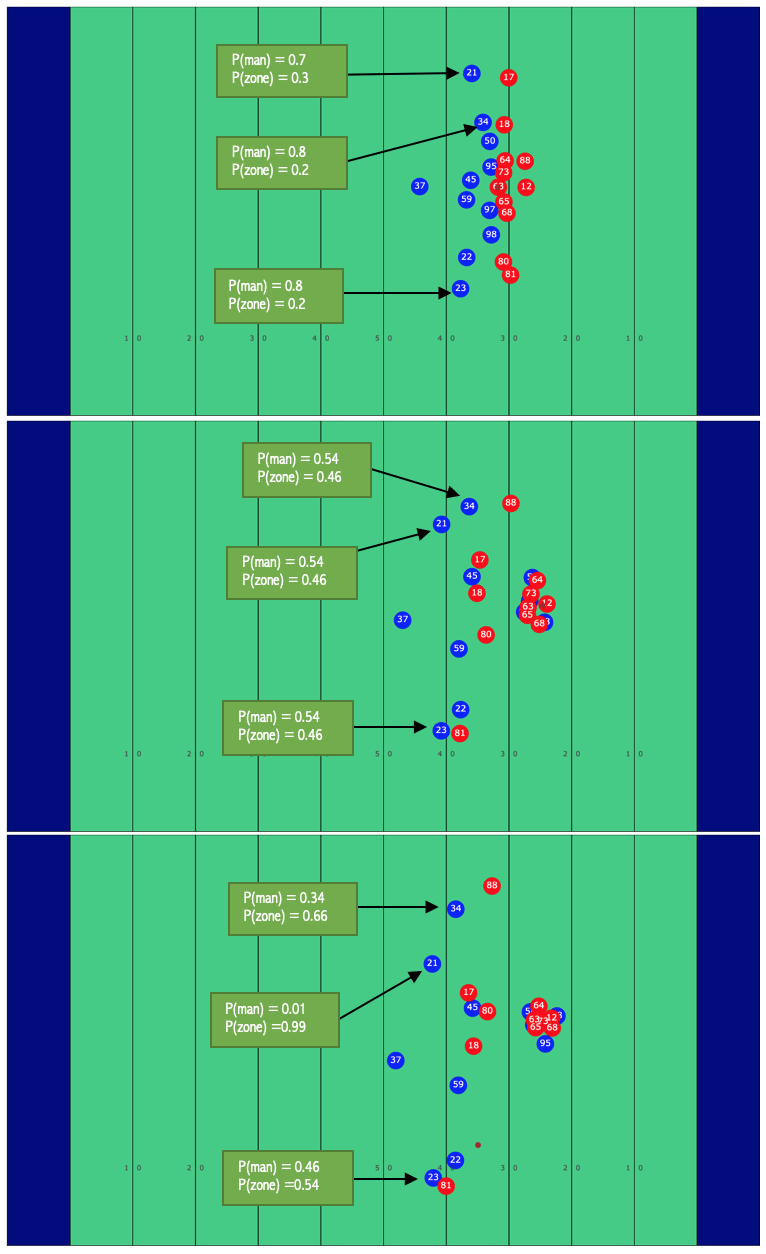}
  \caption{Play 1193 from Game ID 2017091713, with offense shown in red, and defense shown in blue.  The play starts with the formation in the first frame and the ball is thrown in the last frame. The middle frame is exactly in between. In this case the top arrow points to player with jersey 34 (Defense-34), the middle arrow to player with jersey 21 (Defense-21), the bottom arrow points to player with jersey 23 (Defense-23) (in the first frame). In the last frame 21 and 34 swap arrows.}
  \label{fig:play3}
\end{figure}

In Figure \ref{fig:play3}, the offensive is lined up in a ``shotgun'' formation with four wide receivers, two on each side of the offensive line.  Defense-21 and Defense 23 are playing relatively far from their closest receivers, while Defense-34 is relative close to his.  The GMM predicts this starting formation to be man for each of the players, but with some uncertainty in the probabilities ($p \leq 0.8$ for each).  As the play continues, Defense-21 and Defense-34 actually swap positions. In frame 1, Defense-34 is defending Offense-18, but as the play develops, we see that Defense-34 is in man coverage against Offense-88, who runs out from the offensive backfield.  This unusual starting point and pattern of motion by Offense-88 leads to a reduced probability that Defense-34 is in man coverage according to the GMM.  Defense-21 does not follow Offense-17 and instead occupies a zone in the middle frame, where his pattern of motion only slightly follows that of Offense-17.  Finally, Defense-23 does not actually move much from his initial position at the start of the play, but Offense-81 runs a corner route towards his direction.  Interestingly, the model gives about equal probability that Defense-23 is zone (54\%) coverage and man (46\%) coverage, because there is another defensive player nearby who might also be assigned to Offense-81.  His lack of motion and the proximity of another defender is more common with zone coverage, but his proximity to Offense-81 and similar pattern of motion to that of Offense-81 indicates man coverage.  Thus, the model gives a fairly uncertain prediction about his coverage type.  This uncertainty is a positive feature of the mixture model:  We should not assign hard cluster labels when it is unclear, even to the human eye, what type of coverage each player is playing.

\subsection{Analysis of Coverage Types by Pattern of Motion, Player, Team, and Situation}
\label{sec:results-analysis}

\begin{figure}[]
  \centering
  \includegraphics[width=0.9\linewidth]{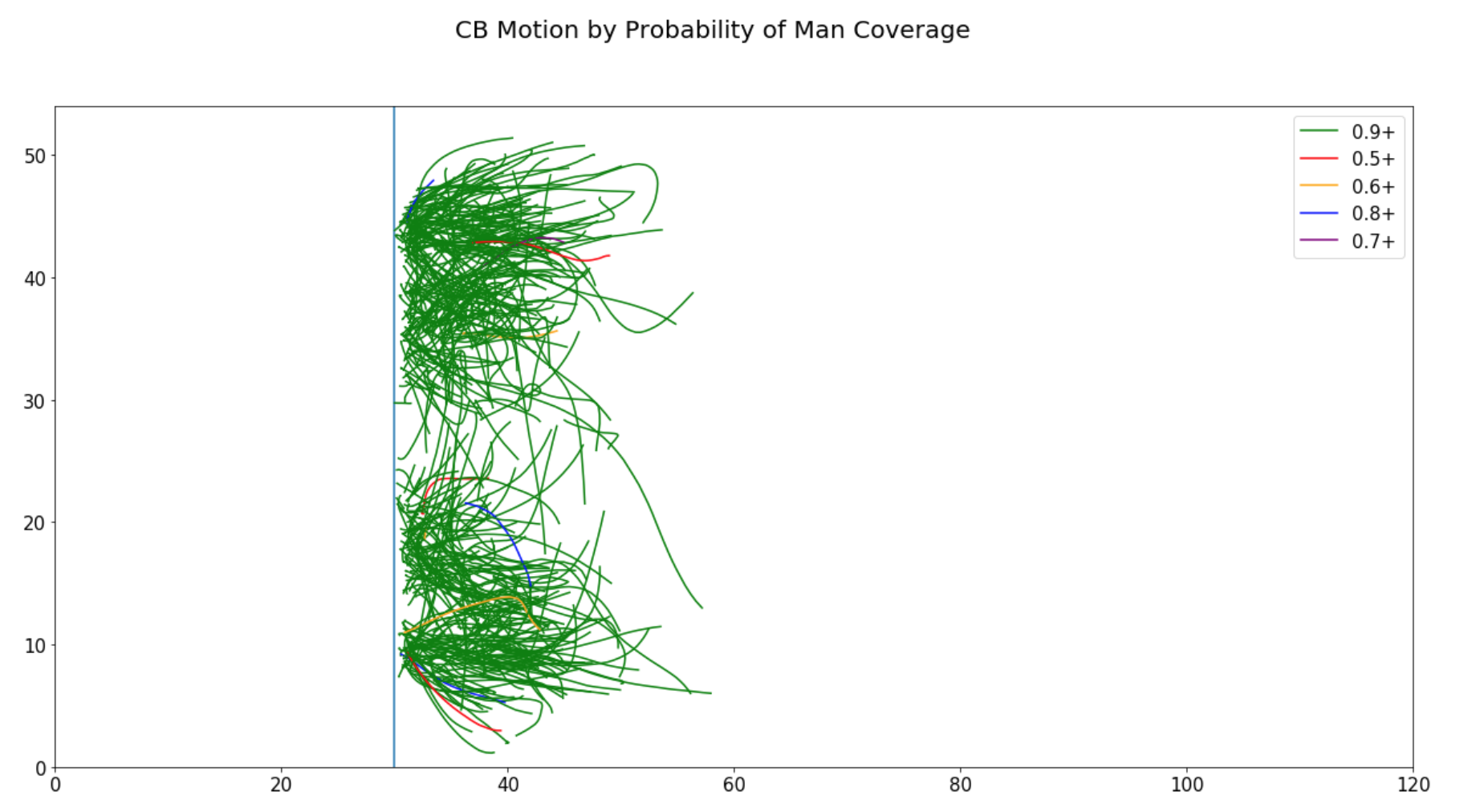}
  \caption{Probability motion is classified as Man}
  \label{fig:sub1}
\end{figure}

\begin{figure}[]
  \centering
  \includegraphics[width=.9\linewidth]{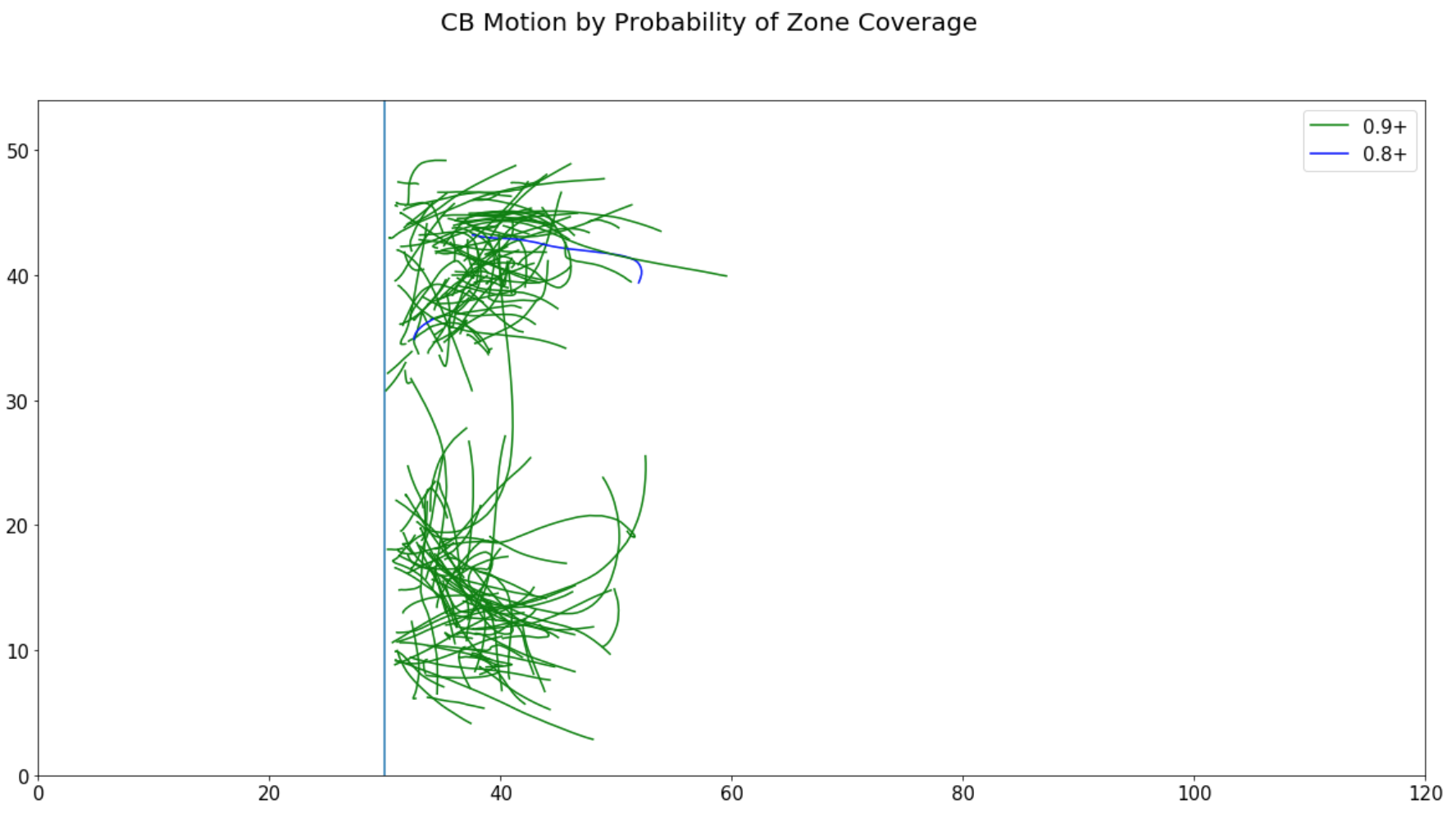}
  \caption{Probability motion is classified as Zone}
  \label{fig:sub2}
\end{figure}

Figures \ref{fig:sub1} and \ref{fig:sub2} show the patterns of motion of cornerbacks classified as playing man or zone coverage, respectively.  Here, we see no apparent relationship between the patterns of motion and the probability of man or zone coverage.  This makes sense in context, since the patterns of motion of cornerbacks are typically reactionary to what the opposing receiver is doing. This provided partial evidence that a trajectory clustering approach like that of \cite{Chu19} or \cite{Miller17} would not be appropriate for identifying coverage types of defensive backs.

%We can see that the GMM predicts man with varying probabilities. Some of the motions, albeit similar, are classified into different categories. This is likely due to the offensive player having different motions, and thus the model was able to capture this distinction. When the GMM predicts zone coverage, the probability is generally higher.

%\FloatBarrier

\section{Discussion and Future Work}
\label{sec:discussion}

We present an unsupervised approach, using Gaussian mixture modeling, for identifying and annotating the type of pass coverage of cornerbacks during passing plays in the NFL, requiring no ground truth labels and minimal human oversight.  We design a rich set of features that help distinguish between types of coverage and can be updated at each point during a play, and find that relate the direction of motion of the cornerback to that of the closest offensive players are most useful when separating ``man coverage'' from ``zone coverage.''  We use an out-of-sample prediction approach and find that a two-cluster solution (corresponding to man coverage and zone coverage) yields the best fit.  We demonstrate how the mixture model's probabilistic cluster assignments allow for interesting subsequent analyses, such as examining how pass coverage types evolve and become more obvious throughout the course of a play.  

We use Gaussian mixture modeling to address this problem, and find that the mixture model's flexibility, interpretability, and soft cluster assignments are more suitable for this problem.  We demonstrate this with several case studies of specific plays, showcasing how our probabilistic coverage type assignments evolve throughout the course of a play in reaction to the on-field movements of cornerbacks in relation to other players on the field.  We show how our model can be used to identify cases when defenses attempt to disguise their coverage types before the snap, before settling into patterns of motion common to the type of coverage they are playing as the play develops.

We use an out-of-sample prediction approach and the well-studied adjusted Rand index to determine the appropriate number of clusters.  Using this approach, we find that the two-cluster solution yields the best ARIs, which conveniently matches our expectations given what we know about the dichotomy of coverage types in football to man and zone coverage.  Furthermore, through manual review of animated example plays from each cluster, we assign ``man'' and ``zone'' labels to the groups identified by the two-cluster GMM solution.

We use a similar approach to determine the utility of each feature in improving the clustering results.  We find that the variability in the direction of motion of the cornerback relative to the nearest offensive player is the most important variable in improving the clustering results, as measured by the adjusted Rand index.  We also find that other features examining the variability in the movement of the cornerback (in relation to offensive players or otherwise) are important in improving the clustering results.  Interestingly, we find that comparing the directions of motion of the cornerback to his nearest teammate does little to improve the clustering results.

The mixture model's probabilistic cluster assignments are advantageous in the context of football, as we demonstrate through an examination of three plays.  In particular, since modern defensive schemes involve disguising coverage types at the start of plays and reacting to what the offense is doing, we examine the GMM cluster membership probabilities at different points throughout the course of a play, showing how these probabilities change in reaction to the patterns of motion of cornerbacks and their proximity to opposing receivers.

We provide a brief analysis of coverage types by different game situations, though we acknowledge that there is more work to be done here.  However, that encapsulates the primary purpose of this research:  to provide additional annotations to NFL player and ball tracking data that will allow future researchers and NFL teams to explore interesting and innovative research problems.  We look forward to seeing such new research.

One limitation of our approach is that player orientation was not available in the data provided by the NFL via the 2019 Big Data Bowl.  This is unfortunate, since this information would be helpful in distinguishing man vs. zone coverage for cornerbacks.  NFL teams have access to this information at the frame-level, and they should be able to easily incorporate this into their feature set when replicating our mixture modeling approach internally.  As an example, we propose a brief set of features involving the players' orientations on the field that may help distinguish man vs. zone coverage.

\begin{enumerate}
    \item \textbf{Orientation of the defensive back relative to the line of scrimmage at different points throughout the play.}  At the start of each play, cornerbacks typically face the line of scrimmage.  As the play develops, cornerbacks in man coverage frequently turn their bodies while following the offensive player (e.g. wide receiver) that they are covering, while cornerbacks in zone coverage more frequently face the line of scrimmage, watching the quarterback's actions as the play develops.  One can design features around this idea.  For example, the percentage of time that a cornerback is facing the line of scrimmage, the direction the cornerback is facing at the time the ball is thrown, etc.
    \item \textbf{Orientation of the cornerback relative to the corresponding offensive player at different points throughout the play.}  As a passing play develops, cornerbacks in man coverage frequently mimic the on-field movements of the offensive player they are covering, so that the direction that they are facing on the field will closely match that of the offensive player.  For cornerbacks in zone coverage, their orientation is more likely to deviate from the closest offensive player.  One can design features around this idea.  For example, one can compute the percentage of time throughout a play in which the two players facing the same direction (e.g. within some angle $\theta < 45$).
\end{enumerate}

These are just examples of how the feature set can be enriched with the addition of player orientation into the dataset.  We encourage those with access to the full NFL player tracking data to explore these options and improve upon our feature set in their own work.

Another limitation of our approach is addressing the correlation between the features considered. We emphasize that this is a first step in understanding the role of our constructed features in modeling defensive coverage, and leave further feature analysis, such as application of other model-based variable selection methods discussed in \citet{Fop2018}, for future work.

In future work, we apply this framework to analyzing the coverage types of safeties and, when appropriate, linebackers.  We have briefly applied the framework to safeties in pass coverage as a proof-of-concept, and found that two coverage type clusters was the best fit, though the size of the clusters were more disparate (most likely, safeties are playing zone coverage more than man coverage).  However, a complete analysis of other positions would require the design of new features specific to the safety position and the patterns of motion of safeties in relation to their teammates and opponents. This task is left to future work.  

Additionally, we hope to examine using the coverage type of one player to help inform the coverage type prediction of another player.  For example, knowing that the left cornerback is playing in man coverage could indicate that the right cornerback is likely also going to play man coverage; knowing that the safeties are playing ``cover-2'' might change our estimate of the probability that the cornerbacks are playing zone coverage.  Additionally, incorporating team-specific features may help, since certain defensive coordinators may prefer certain defensive schemes against certain personnel packages. %The NFL player tracking data also contains player orientation. In order to include this in future work, one could use player orientation as a feature in the clustering and also build additional features that compare a DBs orientation angle with the corresponding WR's angle. We would expect cornerbacks in zone coverage to be angled toward the line of scrimmage more frequently than a man in zone. With reliable player orientation we would likely be able to increase the quality of our model.  

Finally, in future work, we hope to provide additional annotations of other on-field events (actions, coverage schemes, etc).  The mixture modeling approach we use here should serve as a foundation for this future work, although a new set of features will need to be designed for each specific annotation problem. 

\bibliographystyle{DeGruyter}
\bibliography{references}

\section{Appendix}
\begin{figure}[]
  \centering
  \includegraphics[width=.9\linewidth]{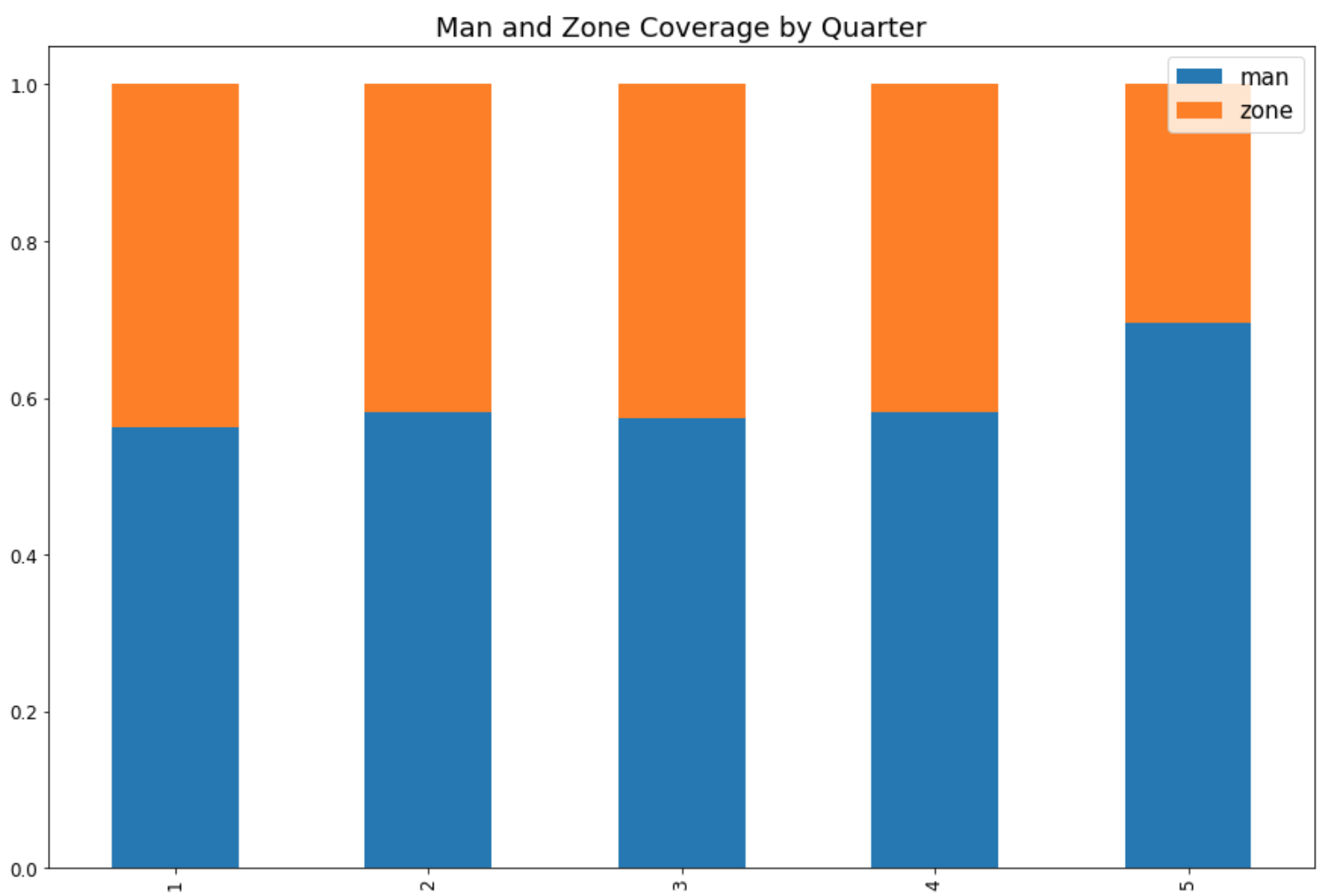}
  \caption{Man:Zone Percentage by Quarter}
  \label{fig:quart}
\end{figure}

Figure \ref{fig:quart} shows the proportion of cornerback coverages in man (blue) and zone (orange) by quarter.  While some small differences exist in the first four quarters, they are mostly negligible, meaning we see no apparent trend in the type of coverage throughout the game.  Interestingly, we see an increase in man coverage in overtime.  However, this result is not statistically significant, since there were only 40 overtime plays in our dataset from the first six weeks of the 2017 season.  This brief analysis does not control for factors like the score differential or opposing offensive formation, which may influence coverage type.  %We leave that exercise to future work.  

\begin{figure}[]
  \centering
  \includegraphics[width=.9\linewidth]{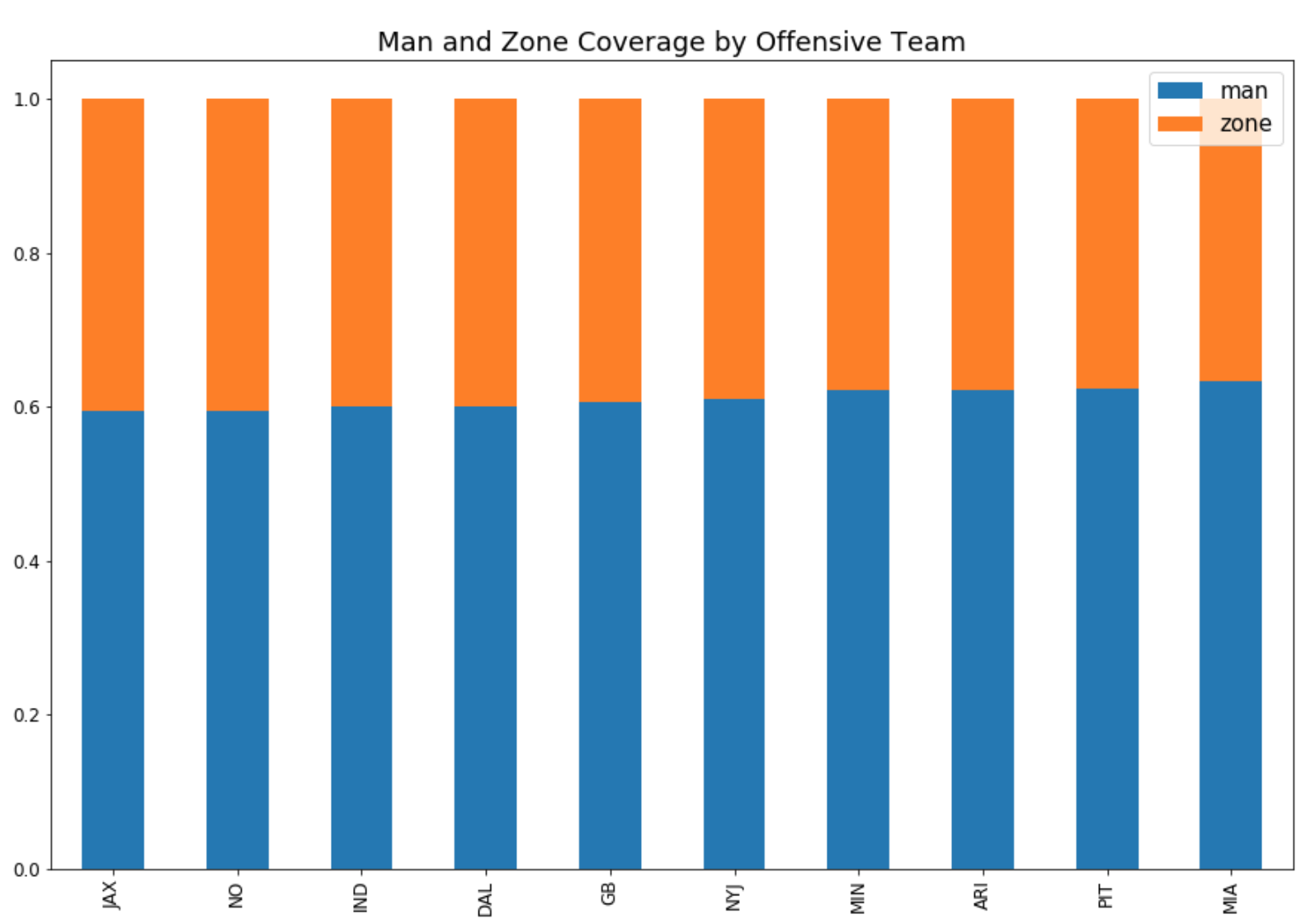}
  \caption{Top 10 teams with the highest Zone Coverage percentage}
  \label{fig:man}
\end{figure}

\begin{figure}[]
  \centering
  \includegraphics[width=.9\linewidth]{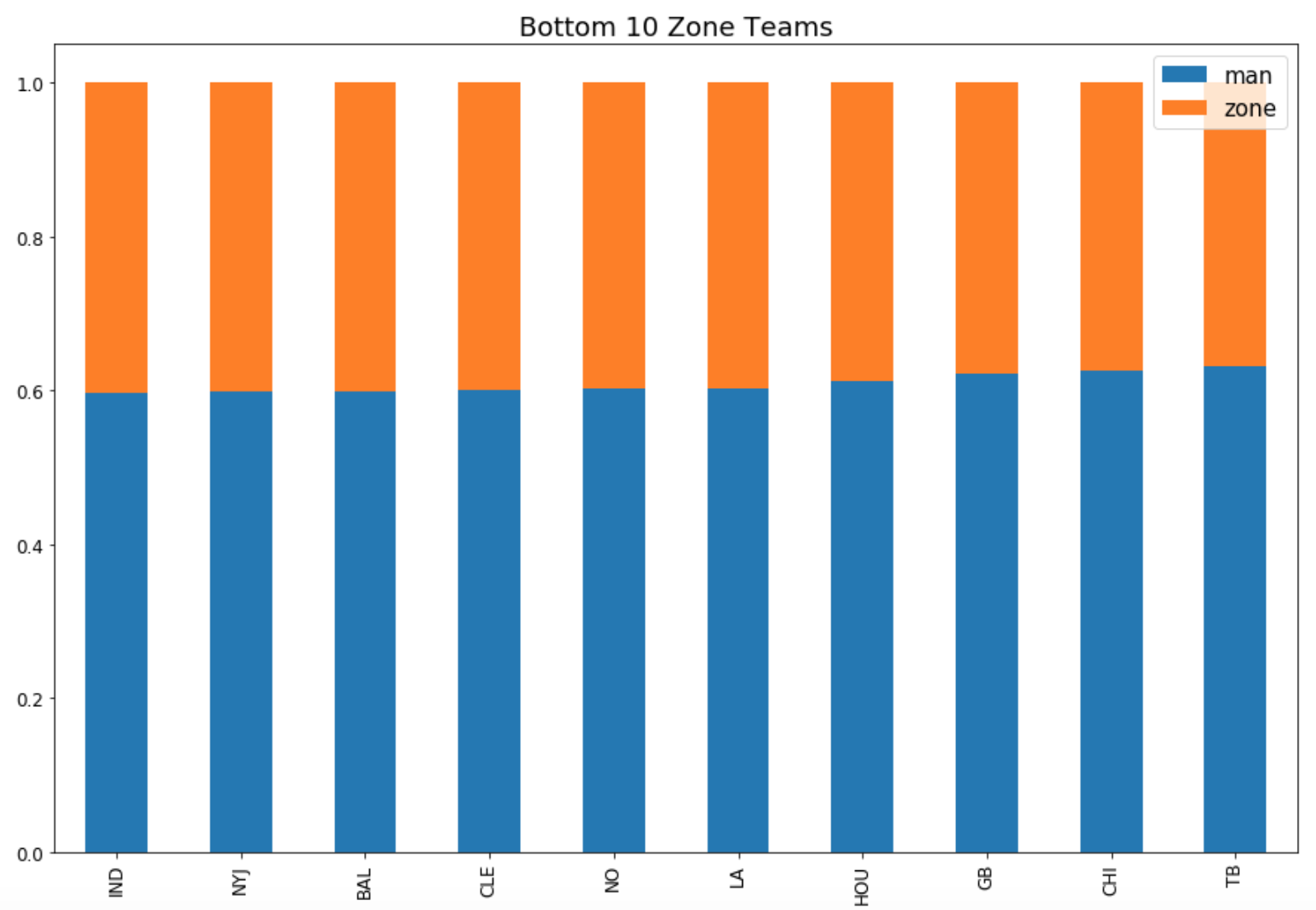}
  \caption{Top 10 teams with the highest Man Coverage percentage}
  \label{fig:zone}
\end{figure}

Figures \ref{fig:man} and \ref{fig:zone} show the top-10 teams by proportion of zone and man coverage, respectively.  Tampa Bay, Chicago, and Green Bay have the highest rate man coverage play during this short span of six weeks, while Washington, Buffalo, and the New York Giants have the highest rate of zone coverage.  Overall, man coverage is used more often than zone coverage by all 32 NFL teams in this provided sample of data.

%We can see from Figure \ref{fig:quart} and Figure \ref{fig:down} that the quarter or down does not impact the use of man or zone coverage. Rather, in the overtime period we see a significant increase in man coverage. By down, we see that man is used more than zone but equally among the downs. Man is increased slightly more during a fourth down pass, but generally is similar to the other downs. 

\begin{figure}[]
  \centering
  \includegraphics[width=.8\linewidth]{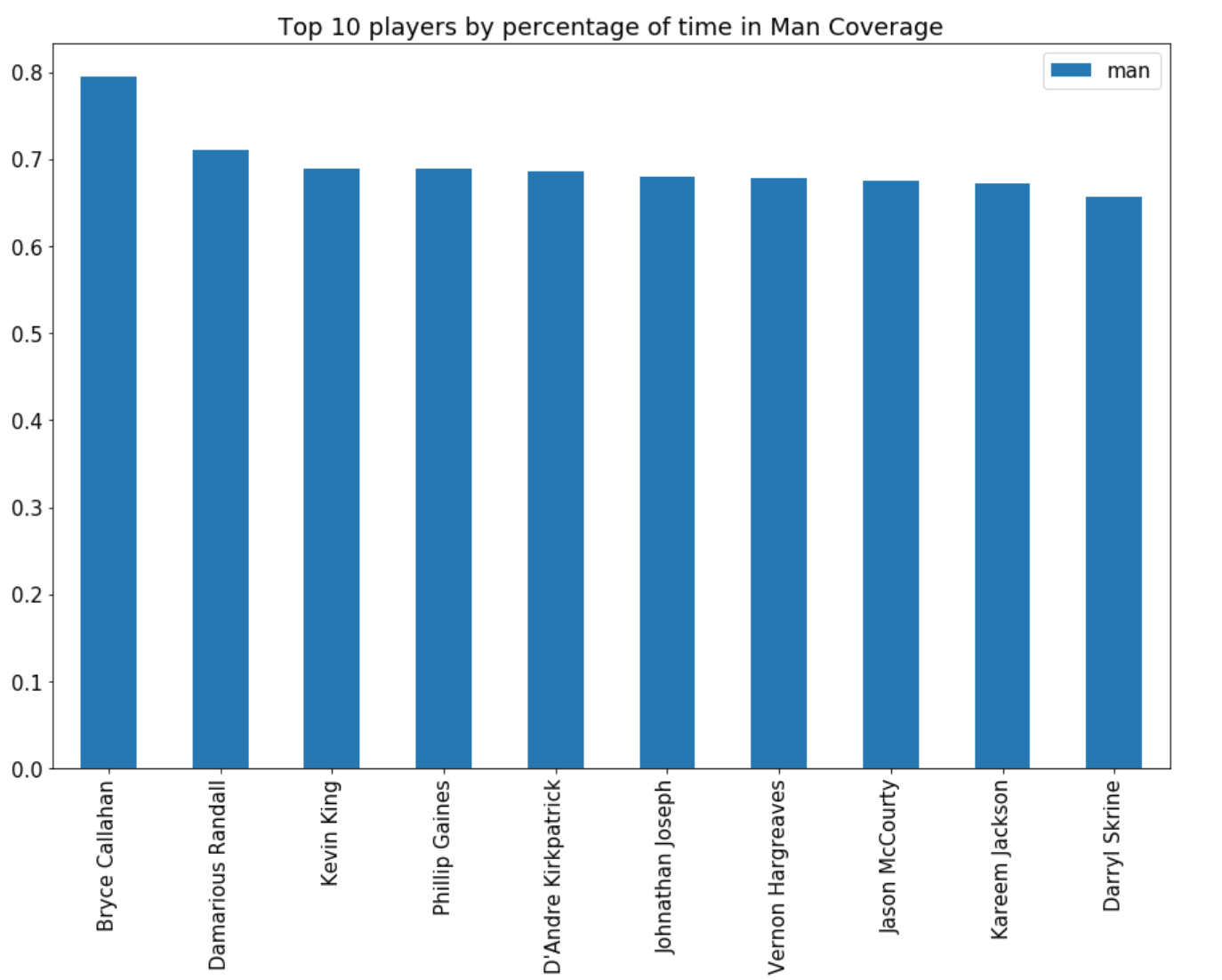}
  \caption{Top 10 players with the highest Man Coverage percentage}
  \label{fig:man_player}
\end{figure}

\begin{figure}[]
  \centering
  \includegraphics[width=.8\linewidth]{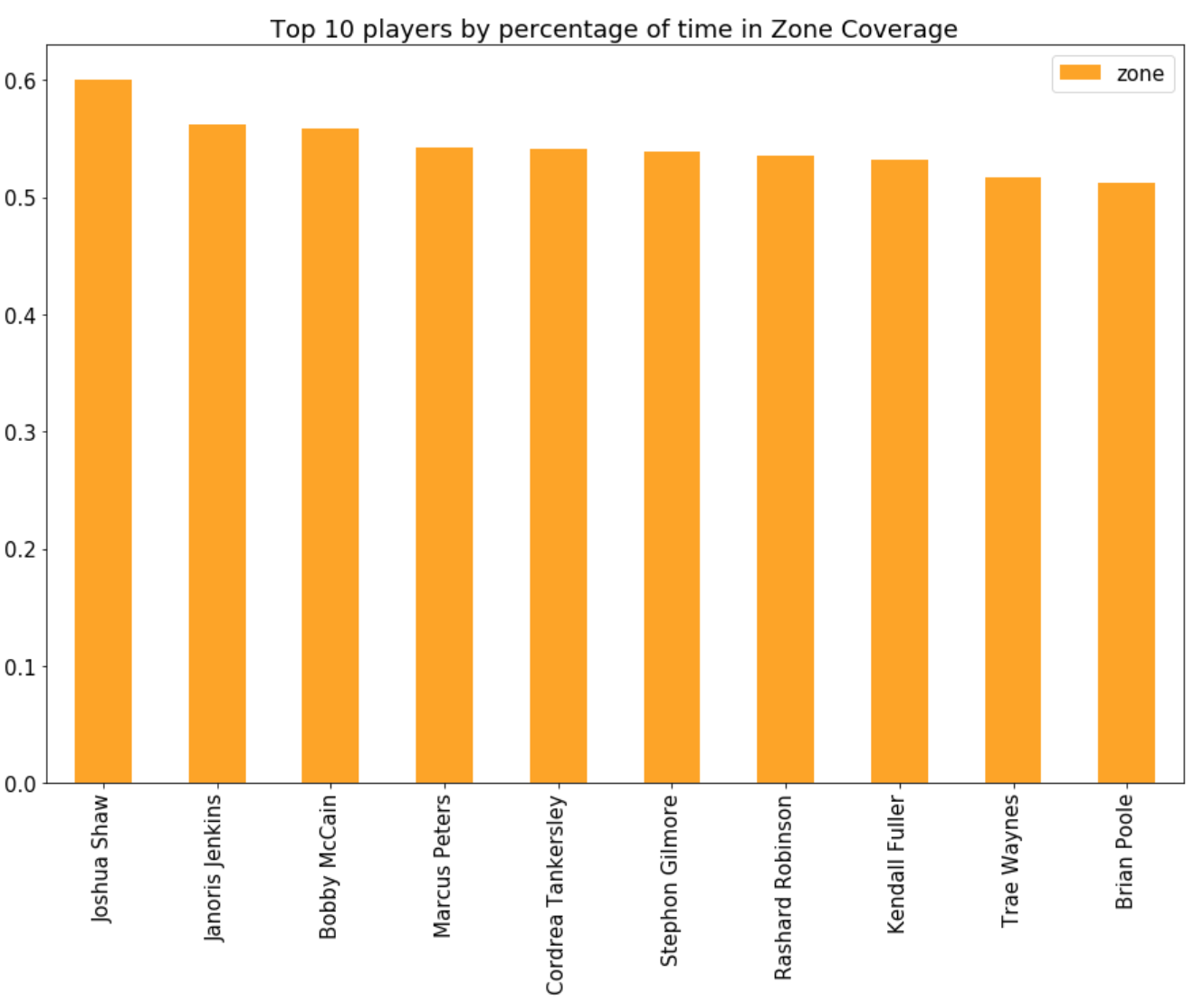}
  \caption{Top 10 players with the highest Zone Coverage percentage}
  \label{fig:zone_player}
\end{figure}

Finally, Figure \ref{fig:man_player} and \ref{fig:zone_player} show the top-10 cornerbacks by their proportion of man and zone coverage, respectively (minimum 50 coverages).  Bryce Callahan from the Chicago Bears led the NFL in percentage of man coverages during the first six weeks of the 2017 season with almost 80\%, followed by Damarious Randall, Kevin King, Phillip Gaines, and others.  Joshua Shaw, who played for the Bengals in 2017, led the league in percentage of zone coverages, with about 60\%, followed by Janoris Jenkins, Bobby McCaine, Marcus Peters, and others.

\begin{figure}[]
  \centering
  \includegraphics[width=.9\linewidth]{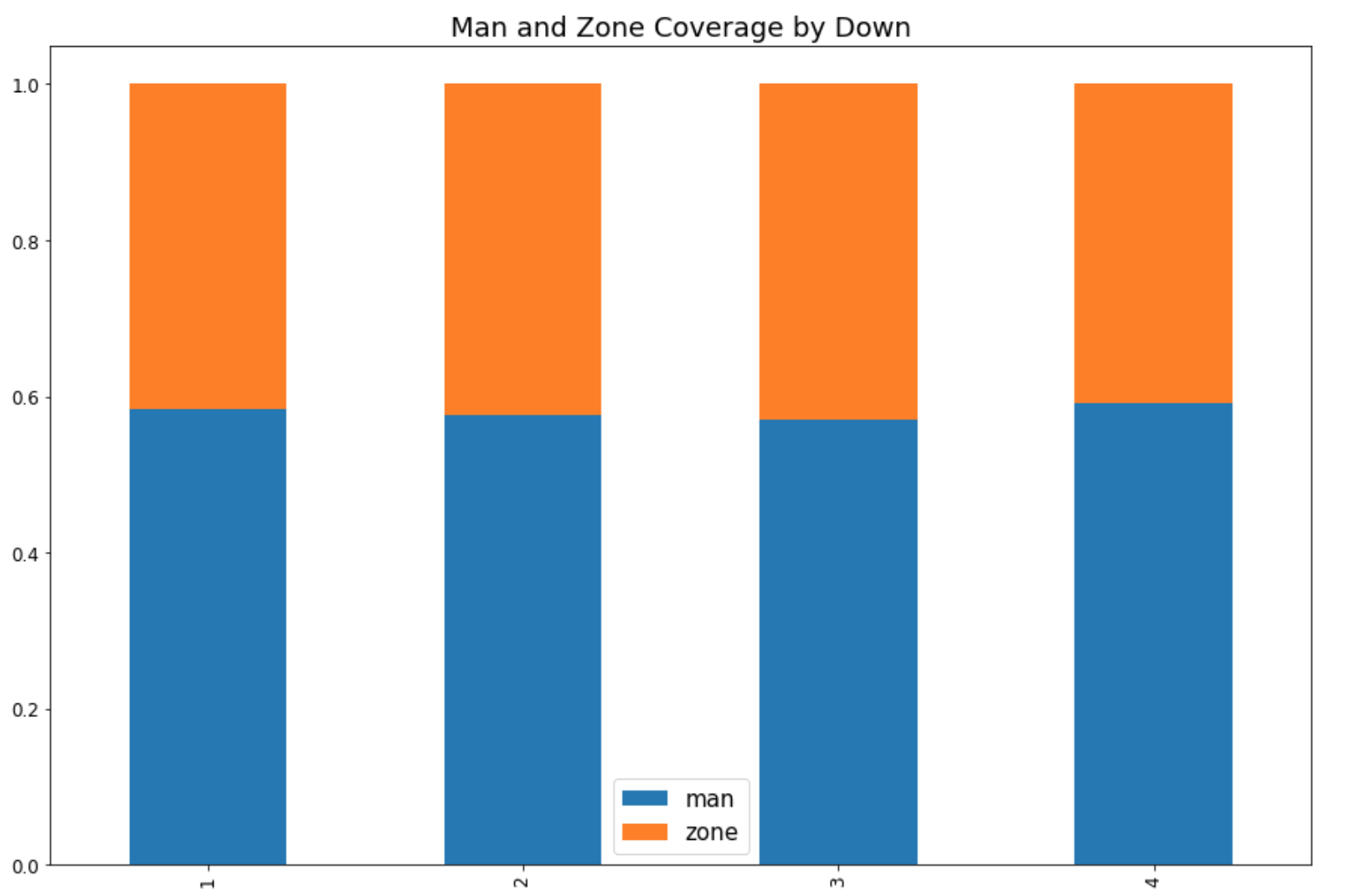}
  \caption{Man:Zone Percentage by Down}
  \label{fig:down}
\end{figure}

Figure \ref{fig:down} shows no apparent relationship between the down and the coverage type of cornerbacks. 

\end{document}